\documentclass[12pt]{amsart}
\usepackage{amsmath,amsthm}
\usepackage{amssymb,mathtools}
\usepackage{etex} 
\usepackage{stmaryrd}
\usepackage{fancyhdr}
\usepackage[normalem]{ulem}
\usepackage{longtable}
\usepackage[nosort]{cite}
\usepackage{calligra} 
\usepackage{dsfont}
\usepackage{enumerate}
\usepackage[all]{xy} 
\input diagxy

\usepackage{url}
\usepackage[colorlinks=true]{hyperref}
\hypersetup{
	colorlinks=true,
    linktoc=all,
    linkcolor=blue,  
    citecolor=red,
    }

\numberwithin{equation}{section}

\theoremstyle{theorem}
\newtheorem{theorem}[equation]{Theorem}
\newtheorem{corollary}[equation]{Corollary}
\newtheorem{proposition}[equation]{Proposition}
\newtheorem{lemma}[equation]{Lemma}
\theoremstyle{definition}
\newtheorem{definition}[equation]{Definition}
\newtheorem{construction}[equation]{Construction}
\newtheorem{example}[equation]{Example}
\newtheorem{remark}[equation]{Remark}
\newtheorem{physics}[equation]{Physics}

\let\a=\alpha \let\b=\beta \let\g=\gamma  \let\e=\epsilon
 \let\h=\eta   
\let\l=\lambda \let\r=\rho
\let\s=\sigma \let\t=\tau  \let\f=\phi  
\let\w=\omega         
  \let\S=\Sigma   
\let\C=\Chi \let\W=\Omega

\newcommand{\be}{\begin{equation}}
\newcommand{\ee}{\end{equation}}
\def\ba{\begin{align}} 
\def\ea{\end{align}}
\newcommand{\bea}{\begin{eqnarray}}
\newcommand{\eea}{\end{eqnarray}}
\newcommand{\bex}{\begin{exercise}}
\newcommand{\eex}{\end{exercise}}
\newcommand{\ban}{\begin{answer}}
\newcommand{\ean}{\end{answer}}
\newcommand{\bt}{\begin{theorem}}
\newcommand{\et}{\end{theorem}}
\newcommand{\bc}{\begin{corollary}}
\newcommand{\ec}{\end{corollary}}
\newcommand{\blem}{\begin{lemma}}
\newcommand{\elem}{\end{lemma}}
\newcommand{\bp}{\begin{problem}}
\newcommand{\ep}{\end{problem}}
\newcommand{\bn}{\begin{proposition}}
\newcommand{\en}{\end{proposition}}
\newcommand{\bd}{\begin{definition}}
\newcommand{\ed}{\end{definition}}
\newcommand{\bcon}{\begin{construction}}
\newcommand{\econ}{\end{construction}}
\newcommand{\bq}{\begin{question}}
\newcommand{\eq}{\end{question}}
\newcommand{\bprf}{\begin{proof}}
\newcommand{\eprf}{\end{proof}}
\newcommand{\br}{\begin{remark}}
\newcommand{\er}{\end{remark}}
\newcommand{\bs}{\begin{solution}}
\newcommand{\es}{\end{solution}}
\newcommand{\beqs}{\begin{eqnarray}}
\newcommand{\eeqs}{\end{eqnarray}}

 \let\ov=\overline

\newcommand{\<}{\langle}
\renewcommand{\>}{\rangle}

\newcommand{\id}{\mathrm{id}}

\newcommand{\mC}{\mathcal{C}}
\newcommand{\mD}{\mathcal{D}}
\newcommand{\up}{\uparrow}
\newcommand{\dn}{\downarrow}

\newcommand{\aand}{\qquad \& \qquad}
\newcommand{\tr}{{\rm tr} }

\def\R{{{\mathbb R}}}
\def\C{{{\mathbb C}}}

\def\Hi{{{\mathcal{H}}}}

\newcommand{\CAlg}{\mathbf{C^*\text{-}Alg}}
\def\mA{{{\mathcal{A}}}}
\def\mS{{{\mathcal{S}}}}
\def\mB{{{\mathcal{B}}}}
\def\mG{{{\mathcal{G}}}}
\newcommand{\Rep}{\mathbf{Rep}}
\newcommand{\Set}{\mathbf{Set}}
\newcommand{\Cat}{\mathbf{Cat}}

\newcommand{\op}{\mathrm{op}}
\newcommand{\St}{\mathbf{States}}
\newcommand{\states}{\scripty{states}}
\newcommand{\GNS}{\mathbf{GNS}}
\newcommand{\rest}{\mathbf{rest}}
\newcommand{\pRep}{\mathbf{Rep}^{\bullet}}
\newcommand{\cRep}{\mathbf{Rep}^{\odot}}
\newcommand{\pGNS}{\mathbf{GNS}^{\bullet}}

\DeclareMathAlphabet{\mathcalligra}{T1}{calligra}{q}{n}
\DeclareFontShape{T1}{calligra}{q}{n}{<->s*[2.2]callig15}{}
\newcommand{\scripty}[1]{\ensuremath{\mathcalligra{#1}}}

\newcommand{\ben}{\renewcommand{\theenumi}{\alph{enumi}} 
\renewcommand{\labelenumi}{(\theenumi)}\begin{enumerate}}
\newcommand{\een}{\end{enumerate}}

\title[From observables and states to Hilbert
space and back]{From observables and states to Hilbert
space and back: a 2-categorical adjunction}
\author[A.~Parzygnat]{Arthur J. Parzygnat}
\address{Mathematics Department, University of Connecticut \\
Storrs, CT 06269, USA\\
{{\em Email:} {\tt\href{mailto:arthur.parzygnat@uconn.edu}
{arthur.parzygnat@uconn.edu}}}}
\begin{document}
\maketitle


\begin{abstract}
Given a representation of a $C^*$-algebra, thought of
as an abstract collection of physical observables, together
with a unit vector, one obtains a state on the
algebra via restriction. 
We show that the Gelfand-Naimark-Segal (GNS) construction
furnishes a left adjoint of this restriction.
To properly formulate this adjoint, it must be
viewed as a weak natural transformation, 
a 1-morphism in a suitable 2-category, 
rather than as a functor between categories.
Weak naturality encodes the functoriality 
and the universal property of adjunctions encodes the
characterizing features of the GNS construction. 
Mathematical definitions and results are accompanied 
by physical interpretations.
\end{abstract} 

\tableofcontents

\section{Introduction and outline}
\label{sec:GNSintro}

There is a familiar construction whose input
consists of a representation of a $C^*$-algebra
on a Hilbert space together with a unit vector
and whose output is a state on
the $C^*$-algebra via restriction. 
Namely, given an algebra $\mA,$ a 
representation $\pi:\mA\to\mB(\Hi)$ to
bounded operators on a Hilbert space $\Hi,$
and a unit vector $\psi\in\Hi,$ one obtains a state
on $\mA$ given by the expectation values of
observables in $\mA$ sending $a\in\mA$ to 
$\<\psi,\pi(a)\psi\>,$ where $\<\;\cdot\;,\;\cdot\;\>$ denotes
the inner product on $\Hi.$ 
We show that this construction,
denoted by $\rest,$ can be expressed
categorically as a natural transformation
\be
\xy0;/r.25pc/:
(-20,0)*+{\CAlg^{\op}}="1";
(20,0)*+{\Cat}="2";
{\ar@/^1.5pc/"1";"2"^{\St}};
{\ar@/_1.5pc/"1";"2"_{\pRep}};
{\ar@{=>}(0,-5);(0,5)_{\rest}};
\endxy.
\ee
Here, $\Cat$ is the category of categories,
$\CAlg$ is the category of $C^*$-algebras,
$\St$ is the functor that associates
a category of states to every $C^*$-algebra,
and $\pRep$ is the functor that associates
the category of representations of $C^*$-algebras
(the $\bullet$ is to denote the additional choice of a unit vector).

Our purpose here is to
prove that the natural transformation
$\rest$ has a left adjoint
\be
\xy0;/r.25pc/:
(-25,0)*+{\CAlg^{\op}}="1";
(25,0)*+{\Cat}="2";
(0,0)*{\dashv};
{\ar@/^1.75pc/"1";"2"^{\St}};
{\ar@/_1.75pc/"1";"2"_{\pRep}};
{\ar@{=>}(2.5,-5.5);(2.5,5.5)_{\rest}};
{\ar@{=>}(-2.5,5.5);(-2.5,-5.5)_{\pGNS}};
\endxy
\ee
denoted by $\pGNS$ because its
ingredients are composed of constructions
due to Gelfand, Naimark, and Segal \cite{GN43},\cite{Se47}, 
which we review.  
We therefore call $\pGNS$ the \emph{GNS construction}. 
This allows us to view the GNS construction as a morphism/process
in some appropriate category rendering it accessible to the
techniques and tools of category theory. 
Although not quite a functor, 
this provides a precise sense in which the GNS construction
is functorial with respect to $*$-homomorphisms of $C^*$-algebras.
Futhermore, the key properties
of the GNS construction are shown to be naturally described 
in terms of categorical concepts 
characterizing it as a left adjoint to the
restriction map from representations
to states. By the essential uniqueness
of adjoints, this offers a 
\emph{definition} of the GNS construction
so that one can view the standard GNS construction
as exhibiting the \emph{existence} of such an adjoint. 

However, there are subtleties in
this description. First, the GNS
construction is a certain
2-categorical natural transformation (utilizing
the fact that $\Cat$ is a 2-category)
instead of a natural transformation
in the usual sense of ordinary category theory \cite{Ma98}.
Second, the category of states
is not the naive one that one might think
of---one must view the states of a fixed $C^*$-algebra
as a discrete category to obtain an appropriate functorial
description. 
Third, for a robust statement with physical
applications, the morphisms in the
representation category associated
to a $C^*$-algebra must include all intertwiners
that are isometries and not only the
unitary equivalences. 

The outline of this paper is as follows. 
Section \ref{sec:states} defines all
relevant notions from $C^*$-algebras
as well as the states functor
and the representation functor.
Section \ref{sec:GNS} describes
the GNS construction as is usually
found in the literature but framed in a
categorical setting. For simplicity,
we ignore the cyclic vector and focus
only on the fact that the GNS construction
produces a representation. In particular,
we prove that the GNS
construction is an oplax-natural
transformation 
(though not a natural transformation)
in Theorem \ref{thm:GNSfunctor}. 
Section \ref{sec:comments} explains
why the category of states
(introduced in Section \ref{sec:states}) 
must have no non-trivial morphisms for our purposes.
Section \ref{sec:adjoint} properly
accounts for the fact that the GNS
construction produces a \emph{cyclic}
representation. The statement
that the GNS construction
is left adjoint to the restriction
to states natural transformation
is proved in Theorem \ref{thm:GNSmain}.
Definition \ref{defn:GNS} gives a categorical definition
for the GNS construction motivated by our results.
A summary of our results characterizing the GNS construction
is given after this definition. 
In Section \ref{sec:GNSex}, we illustrate
several of the constructions and results
in terms of a simple example of a bipartite 
system familiar (to physicists) from the 
Einstein-Podolsky-Rosen (EPR) setup. 
Throughout, we provide physical
interpretations of most definitions, constructions, and results, 
though some interpretations are heuristic 
rather than rigorous.
Although we assume the
reader is familiar with some basics of
category theory \cite{Ma98}, we include a short 
appendix on 2-categories and
2-categorical adjunctions. 
Otherwise, we aim to be mostly self-contained.

\section{States and representations of $C^*$-algebras}
\label{sec:states}

For more details on $C^*$-algebras,
the reader is referred to \cite{Di77} and \cite{Fi96}. 

\bd
A \emph{\uline{unital Banach algebra}}
is a vector space $\mA$ together with 
\begin{enumerate}[i)]
\item
a binary 
multiplication operation
$\mA\times\mA\to\mA,$
\item
a norm 
$\lVert \ \cdot \ \rVert:\mA\to\R_{\ge0},$ and
\item
an element $1_{\mA}\in\mA.$ 
\end{enumerate}
The multiplication must be distributive
over vector addition, the scalar multiplication must satisfy
$k(ab)=(ka)b=a(kb)$ for all $k\in\C$ and $a,b\in\mA,$ 
the element $1_{\mA}$ must satisfy
$a1_{\mA}=1_{\mA}a=a$ for all $a\in\mA,$
and finally, 
all Cauchy sequences must converge.
\ed

\bd
\label{defn:C*algebra}
A \emph{\uline{unital $C^*$-algebra}}
is a unital Banach algebra 
$\mA$ 
with an involution ${}^{*}:\mA\to\mA$ that is an
anti-homomorphism for the multiplication
and satisfies $\lVert aa^*\rVert=\lVert a\rVert^2$
for all $a\in\mA.$ 
An element $a\in\mA$ is 
\emph{\uline{self-adjoint}} if $a^*=a,$
an \emph{\uline{isometry}} if $a^*a=1_{\mA},$ and
\emph{\uline{unitary}} if $a^*a=1_{\mA}=aa^*.$
\ed

\bd
\label{defn:c*algebramorphism}
Let $\mA$ and $\mA'$ be two unital $C^*$-algebras. 
A \emph{\uline{map/morphism of unital $C^*$-algebras}}
from $\mA'$ to $\mA$ is a bounded linear map
$f: \mA'\to\mA$  
such that $f(a'^*)=f(a')^*,$ $f(a'_{1}a'_{2})=f(a'_1)f(a'_2),$
and $f(1_{\mA'})=1_{\mA}$ for all $a',a'_1,a'_2\in\mA'.$
\ed

\bd
\label{defn:CAlg}
Let $\CAlg$ be the category of unital $C^*$-algebras. Namely,
the objects of $\CAlg$ are unital $C^*$-algebras and the morphisms
are maps of unital $C^*$-algebras. 
\ed

Throughout this article, \emph{all} $C^*$-algebras and their morphisms
will be assumed unital and we will avoid
overuse of this adjective unless it is necessary to stress it.

\bd
\label{defn:state}
Given a $C^*$-algebra $\mA,$ 
a \emph{\uline{state}} on $\mA$ is a
bounded 
linear function ${\w:\mA\to\C}$ such that
$\w(1_{\mA})=1$ and $\w(a^*a)\ge0$ for all $a\in\mA.$
Denote the set of states 
on a $C^*$-algebra $\mA$
by $\mS(\mA).$%
\footnote{$\mS(\mA)$ is more than just a set---it is a convex set,
though this is irrelevant for our present discussion.}
\ed

\bd
\label{defn:RepA}
Let $\Rep(\mA)$ be the category of
representations of the $C^*$-algebra $\mA$
on Hilbert spaces. This means the objects
are pairs $(\pi,\Hi)$ with $\Hi$ a Hilbert space
and $\pi:\mA\to\mB(\Hi)$ a map of
$C^*$-algebras. Here $\mB(\Hi)$ is the algebra of bounded
operators on $\Hi$---the involution on
$\mB(\Hi)$ is taking the Hilbert space adjoint. Morphisms
$(\pi,\Hi)\to(\pi',\Hi')$ are intertwiners, 
i.e. bounded linear operators 
$L:\Hi\to\Hi'$ such that
\be
\label{eq:intertwiner}
L\circ\pi(a)=\pi'(a)\circ L
\quad
\text{ for all $a\in\mA.$}
\ee
\ed

\br
It is very important that we assume the
morphisms in $\Rep(\mA)$ are 
intertwiners and not necessarily just
unitary equivalences. 
This will allow for a wider range of operations that occur in
physics as will be explained later.
\er

\begin{physics}
We think of a $C^*$-algebra $\mA$ as the
algebra of observables of a physical system.%
\footnote{Actually, $\mA$ contains
un-observable operators because it
contains elements that are not
self-adjoint. Examples include
creation and annihilation operators.
In fact, it contains observables that are
self-adjoint but need not be things we
can actually measure in a lab
(such as momentum to the $8^{\text{th}}$
power).
Nevertheless, we call $\mA$ the algebra
of observables by slight abuse of
terminology.}
An
example to relate to is the case
$\mA=\mB(\Hi)$ of bounded operators
on a Hilbert space $\Hi.$
However, the main point of this abstract
perspective is to place the emphasis
on the observables rather than the Hilbert space
of vectors or the particular realization
of an abstract observable as an operator.
For example, we can think of angular momentum
being defined in different ways on different
Hilbert spaces (or even classically on
phase space), but when we think of
angular momentum, we do not think
of which Hilbert space it acts on---we
just think angular momentum!

Furthermore, we do not measure vectors
in a Hilbert space. What we measure
are expectation values. This is precisely
the meaning of a state
$\w:\mA\to\C$ as defined above. A
state assigns an expectation value to
each physical observable. That is what a
physical state is: a list of expectation
values for all our observables (satisfying
reasonable postulates).
For instance, if $a$ is self-adjoint,
then $\w(a)$ is the expectation value of
$a$ and $\w(a^2)-(\w(a))^2$ is the variance.
Therefore, the definition of state includes not only
expectation values of observables, but also
their moments. 

Of course, technically
thinking of observables as an
algebra is an idealization because
observables (as described by the working
physicist) are not always bounded
operators and therefore they do not
form an algebra in the strict sense \cite{HaQM}. 
We will ignore this issue and consider
observables that correspond to bounded
operators.
\end{physics}

The above definitions of $\mS(\mA)$
and $\Rep(\mA)$ extend to functors.

\bn
\label{prop:statefunctor}
The assignment
sending a $C^*$-algebra $\mA$ to $\mS(\mA)$ 
and sending a morphism $\mA'\xrightarrow{f}\mA$
of $C^*$-algebras to $\mS(\mA')\xleftarrow{\mS(f)}\mS(\mA),$
where $\mS(f)$ is defined by
\be
\mS(\mA)\ni\w\mapsto\w\circ f\in\mS(\mA'),
\ee
defines a functor%
\footnote{For any category $\mathcal{C},$
the \emph{opposite category} $\mathcal{C}^{\op}$
has the same objects as $\mathcal{C}$ 
but a morphism from an object $a$
to an object $b$ in $\mathcal{C}^{\op}$
is a morphism from $b$ to $a$ in
$\mathcal{C}.$
Also, $\Set$ is the category of sets.
}
\be
\CAlg^{\op}\xrightarrow{\mS}\Set,
\ee
henceforth referred to as the 
\emph{\uline{states pre-sheaf}}.
\en

\bprf
First, $\w\circ f$ is a state on $\mA'$ because $\w$ and $f$ are linear, 
\be
\w\big(f(1_{\mA'})\big)=\w(1_{\mA})=1,
\ee
and
\be
\w\big(f(a'^*a')\big)=\w\big(f(a')^*f(a')\big)\ge0
\ee
for all $a'\in\mA'.$ $\mS$ is functorial
because the identity $\id_{\mA}:\mA\to\mA$
gets sent to the identity and the composition
of $C^*$-algebra maps 
$\mA''\xrightarrow{f'}\mA'\xrightarrow{f}\mA$ gets
sent to $\mS(f\circ f')=\mS(f')\circ\mS(f).$%
\footnote{The flipping of the order of morphism
composition in the equation 
$\mS(f\circ f')=\mS(f')\circ\mS(f)$ 
is why we use $\op$ in $\CAlg^{\op}.$
This is sometimes referred to as an anti-homomorphism property
or contravariance as opposed to covariance. 
}
\eprf

\begin{physics}
\label{phys:macrostates}
The meaning of this functor physically can
be understood by considering a special case, which
will be used throughout this work.
Suppose $\mA_{0}$ is a subalgebra of
physical observables of $\mA.$ Let
$i:\mA_{0}\hookrightarrow\mA$ be the
inclusion map. The functor $\mS(i)$
takes a state $\w:\mA\to\C$
that gave expectation values
for all observables in $\mA$
and it \emph{restricts} that state
to only give expectation values for a smaller
collection of observables, mathematically
described by $\mA_{0}.$ In thermodynamic
or statistical-mechanical terminology,
one can imagine $\mA$ as describing the algebra
of observables for \emph{microstates}
and $\mA_{0}$ as
describing the set of observables for
some \emph{macrostates}.%
\footnote{I would like to thank V. P. Nair for discussions
on these points.}
In fact, Jaynes used a closely related idea,
that is actually more physically reasonable,
by assuming that $\mA_{0}$ is just a
sub\emph{set} of $\mA$ and develops 
thermodynamics from it \cite{Ja1}.
In this process of restricting to a subalgebra, one therefore loses some
information about the state---we only
know fewer of its expectation values. 
Note that focusing on subalgebras in explaining the physics is not 
too restrictive since, for example, every (unital) $C^*$-algebra map
of finite-dimensional matrix algebras is unitarily equivalent to one of the form 
\be
A\mapsto\begin{bmatrix}A&&\\&\ddots&\\&&A\end{bmatrix}
\ee
for all matrices $A$ in the source of the $C^*$-algebra map \cite{Fi96}. 
Here, all the empty positions are filled with $0$'s. 
Such maps show up whenever there is a local decomposition of
a Hilbert space into tensor products and an observer only has access
to one of these components (see 
Section \ref{sec:GNSex} for more details). 
\end{physics}

There is a functor $\mathcal{D}:\Set\to\Cat$
from the category of sets to the category of
categories given by sending a set to the discrete
category with only identity morphisms. 
More explicitly, a set $X$ gets sent to the category $\mD(X)$
whose set of objects is $X$ and whose set of morphisms
consists only of identities. A function $f:X\to Y$ of sets gets sent to the
functor $\mD(f):\mD(X)\to\mD(Y)$ whose value on objects agrees
with $f.$ This determines $\mD(f).$ 
Thus,
since the composition of functors is a functor,
this gives a functor 
\be
\label{eq:statesfunctor}
\CAlg^{\op}\xrightarrow{\mS}\Set
\xrightarrow{\mathcal{D}}\Cat,
\ee
which we denote by $\St$ and call it
the \emph{\uline{states pre-stack}}.
The categorically-minded reader will immediately
point out that $\Cat$ is actually a 2-category,
and we will indeed use this fact in a
crucial way when we describe the GNS
construction. But for now, let us put this aside.

\bn
The assignment%
\footnote{In the second line of (\ref{eq:Repstack}), the assignment
on objects is described. In the third line, the assignment
on morphisms is specified. We will often use this notation to specify
functors.}
\be
\label{eq:Repstack}
\begin{split}
\CAlg^{\op}&\xrightarrow{\Rep}\Cat\\
\mA&\mapsto\Rep(\mA)\\
\Big(\mA'\xrightarrow{f}\mA\Big)
&\mapsto
\Big(\Rep(\mA')\xleftarrow{\Rep(f)}\Rep(\mA)\Big),
\end{split}
\ee
is a functor.
Here $\Rep(f),$ sometimes written as $f^*,$
is the functor defined by sending
a representation $(\pi:\mA\to\mB(\Hi),\Hi)$
to the representation $(\pi\circ f:\mA'\to\mB(\Hi),\Hi)$
and by sending an intertwiner
$(\pi,\Hi)\xrightarrow{L}(\rho,\mathcal{V})$
to the intertwiner 
$(\pi\circ f,\Hi)\xrightarrow{L}(\rho\circ f,\mathcal{V}).$%
\footnote{The same notation $L$ 
is used because it is
the same operator $L:\Hi\to\mathcal{V}$
at the level of Hilbert spaces.}
$\Rep$ is also called the 
\emph{\uline{representation pre-stack}}.
\en

\bprf
Let us first make sure
$\Rep(f)$ itself is indeed a functor. For $L$
to be an intertwiner in $\Rep(\mA'),$ it must
be that
\be
L\circ\pi\Big(f(a')\Big)=\rho\Big(f(a')\Big)\circ L
\ee
for all $a'\in\mA'.$ This is true because
$f(a')\in\mA$ and 
$L$ is an intertwiner in $\Rep(\mA).$
It is not difficult to see that
$\id_{\mA}$ gets sent to $\id_{\Rep(\mA)}$
and the composition of 
$\mA''\xrightarrow{f'}\mA'\xrightarrow{f}\mA$
gets sent to $\Rep(f')\circ\Rep(f).$
\eprf

\begin{physics}
The meaning of the functor 
(\ref{eq:Repstack}) is as follows.
With each abstract algebra of observables,
there is a collection of Hilbert spaces on
which we can realize these observables.
This collection is not just a \emph{set}%
\footnote{Technically, it is not even a set in
the strict sense,
but that is not the point we are trying to make.}
but a \emph{category} because
there are intertwiners between representations.
Intertwiners are ubiquitous in physics. 
Every tensor operator is
an intertwiner. For instance,
the angular momentum for
particles in three-dimensional space
is a \emph{vector} of operators. This
vector of operators is precisely an
intertwiner \cite{Ha15}. Other
examples of intertwiners
are unitary equivalences of representations.
These are (some of the) 
symmetries of quantum mechanics.
For instance, different observers might
associate a slightly different Hilbert space
to a collection of observables.
In particular, the observables themselves
might be expressed differently. The
position and momentum representations
of basic quantum mechanics provide one
example. The unitary
map defined by the Fourier transform
is an intertwiner (a unitary equivalence)
of representations.
The category of representations conveniently 
packages all of these structures together
in a single mathematical entity. 
\end{physics}

\section{The GNS construction: from observables
and states to Hilbert spaces}
\label{sec:GNS}

We will split the GNS construction
 into three parts.
First, we will describe the construction as is common
in the literature. Then we will describe something
that is less commonly illustrated, and is described
nicely for physicists in \cite{BGdQRL}, which is what the 
GNS construction gives
for $C^*$-algebra morphisms
(and not necessarily just 
$C^*$-algebra {\emph{iso}morphisms}).
The GNS construction was first introduced
by Segal in \cite{Se47} and we will utilize many
of the facts proved in that work. 
At the end of this section, 
we state our first result, Theorem \ref{thm:GNSfunctor}, which says
that the GNS construction is an oplax-natural
transformation 
(see Definition 
\ref{defn:semipseudonaturaltransformation} in
the Appendix)
between the functors
introduced in the previous section.

\bcon
\label{con:GNSobjects}
Let $\w:\mA\to\C$ be a state on a unital $C^*$-algebra $\mA.$
Then the function
\be
\label{eq:GNSpreinnerproduct}
\begin{split}
\mA \times \mA &\to \C\\
(b,a)&\mapsto\w(b^*a)
\end{split}
\ee
is a map that is conjugate-linear in its first variable 
and linear in its second.
Furthermore, it satisfies%
\footnote{Proof: By assumption 
$\w\big((\a a+\b b)^*(\a a+\b b)\big)\ge0$ for all $\a,\b\in\C$
and $a,b\in\mA,$ which in particular
implies that $\w\big((\a a+\b b)^*(\a a+\b b)\big)$ is real.
Equating this expression with its conjugate gives
$\ov\a\b\w(a^*b)+\a\ov\b\w(b^*a)
=\a\ov\b\overline{\w(a^*b)}+\ov\a\b\overline{\w(b^*a)}.$
Setting $\a=\sqrt{-1}$ and $\b=1$ gives
$-\w(a^*b)+\w(b^*a)=\overline{\w(a^*b)}-\overline{\w(b^*a)}$
while setting $\a=1$ and $\b=1$ gives
$\w(a^*b)+\w(b^*a)=\overline{\w(a^*b)}+\overline{\w(b^*a)}.$
Adding these two gives
$2\w(b^*a)=2\overline{\w(a^*b)},$ which proves the claim.
}
\be
\label{eq:states1}
\w(b^*a)=\overline{\w(a^*b)} \qquad \forall \ a,b\in\mA
\ee
and%
\footnote{Proof (this is more or less a standard proof of the 
Cauchy-Schwarz inequality): This splits up into two cases.
First, if $\w(b^*a)=0,$ then the claim is true.
In the other case, suppose that $\w(b^*a)\ne0.$
As in the previous footnote, 
consider the inequality
$\w\big((\a a+\b b)^*(\a a+\b b)\big)\ge0$ valid 
for all $\a,\b\in\C$ and $a,b\in\mA.$ Choose
$\a=\frac{|\w(b^*a)|}{\w(b^*a)}\sqrt{\w(b^*b)}$
and
$\b=-\sqrt{\w(a^*a)}.$ Then, 
$\w\big((\a a+\b b)^*(\a a+\b b)\big)
=2\w(b^*b)\w(a^*a)-2|\w(b^*a)|\sqrt{\w(b^*b)\w(a^*a)}$
using (\ref{eq:states1}) along the way to cancel some terms.
Rearranging and canceling the factor of $2$ gives
$|\w(b^*a)|\sqrt{\w(b^*b)\w(a^*a)}\le\w(b^*b)\w(a^*a).$
Squaring both sides and canceling the common terms
proves the claim.
}
\be
\label{eq:states2}
\left|\w(b^*a)\right|^2\le\w(b^*b)\w(a^*a) \qquad \forall \ a,b\in\mA.
\ee
Define the set of \emph{\uline{null vectors}} by
\be
\label{eq:nullspace}
\mathcal{N}_{\w} := \{x\in\mA \;:\; \w(x^*x)=0\}.
\ee
From (\ref{eq:states2}), it follows that 
\be
\label{eq:stateideal}
|\w(a^*x)|^2\le\w(a^*a)\underbrace{\w(x^*x)}_{0}=0 
\; \implies\; \w(a^*x)=0 
\quad\forall\; a\in\mA, \;x\in\mathcal{N}_{\w}.
\ee
Using this fact,
\be
\label{eq:idealadditive}
\w\big((x+y)^*(x+y)\big)=
\w(x^*x)+\w(y^*y)+\w(x^*y)+\w(y^*x)
=0
\quad\forall\;x,y\in\mathcal{N}_{\w}
\ee
and 
\be
\label{eq:ideal}
\w\big((ax)^*(ax)\big)=
\w(x^*a^*ax)
=\w\big((a^*ax)^*x\big)=0
\quad\forall\;a\in\mA,\;x\in\mathcal{N}_{\w},
\ee
which together show that $\mathcal{N}_{\w}$ is a left ideal 
inside $\mA,$ meaning that $\mathcal{N}_{\w}$ is 
an additive subgroup of $\mathcal{A}$ and $ax\in\mathcal{N}_{\w}$
whenever $a\in\mA$ and $x\in\mathcal{N}_{\w}.$
Furthermore, note that 
(\ref{eq:states1}) and (\ref{eq:stateideal}) imply
\be
\label{eq:stateideal2}
\w(x^*a)=0\quad\forall\;x\in\mathcal{N}_{\w},\;a\in\mA.
\ee
Now, write the equivalence class of $a\in\mA$ 
in the quotient vector space $\mA/\mathcal{N}_{\w}$ as $[a].$
The function (\ref{eq:GNSpreinnerproduct}) descends to a well-defined
inner product
\be
\label{eq:GNSinnerproduct}
\begin{split}
\mA/\mathcal{N}_{\w}\times\mA/\mathcal{N}_{\w}
&\xrightarrow{\< \ \cdot \ , \ \cdot \ \>_{\w}} \C\\
([b],[a])&\mapsto\w(b^*a)
\end{split}
\ee
on $\mA/\mathcal{N}_{\w}$ 
by choosing representatives of the equivalence classes. 
$\< \;\cdot\;,\;\cdot\;\>_{\w}$ is well-defined because for
any other representatives $b'$ and $a'$ of
$[b]$ and $[a],$ respectively, so
that $b-b',a-a'\in\mathcal{N}_{\w},$ 
\be
\begin{split}
\w(b'^*a')&=\w\Big(\big(b-(b-b')\big)^*\big(a-(a-a')\big)\Big)\\
&=\w(b^*a)-\w\Big((b-b')^*a\Big)
-\w\Big(b^*(a-a')\Big)
+\w\Big((b-b')^*(a-a')\Big)\\
&=\w(b^*a)-0-0+0 
\quad \mbox{ by (\ref{eq:stateideal}) and (\ref{eq:stateideal2})} \\
&=\w(b^*a).
\end{split}
\ee
$\< \;\cdot\;,\;\cdot\;\>_{\w}$ is positive definite 
by definition of $\mA/\mathcal{N}_{\w}$ and because $\w$ is a state. 
Complete $\mA/\mathcal{N}_{\w}$ with respect
to the norm $\lVert\;\cdot\;\rVert_{\w}$ 
induced by $\<\;\cdot\;,\;\cdot\;\>_{\w}$
and denote this Hilbert space by
\be
\label{eq:GNSHilbertspace}
\Hi_{\w}:=\overline{\mA/\mathcal{N}_{\w}}.
\ee
There is a natural action $\pi_{\w}$ of $\mA$ on 
$\mA/\mathcal{N}_{\w}$ given by%
\footnote{This is well-defined because 
$\mathcal{N}_{\w}$ is a left ideal in $\mA$ 
by (\ref{eq:idealadditive}) and (\ref{eq:ideal}).
}
\be
\label{eq:GNSrepn}
\pi_{\w}(a) [b] := [ab]
\ee
for all $a\in\mA$ and $[b]\in\mA/\mathcal{N}_{\w}.$
$\pi_{\w}(a)$ is a bounded operator on $\mA/\mathcal{N}_{\w}$
for all $a\in\mA$ because%
\footnote{In the second last step, we have used the fact
that $\w(y^*xy)\le\lVert x\rVert\w(y^*y)$ for all $x,y\in\mA$
(see Proposition 2.1.5. part (ii) of \cite{Di77} for a proof).}
\be
\lVert\pi_{\w}(a)[b]\rVert_{\w}^2
=\lVert[ab]\rVert_{\w}^2
=\w\big((ab)^*ab\big)
=\w(b^*a^*ab)
\le\lVert a^*a\rVert\w(b^*b)
=\lVert a^*a\rVert\lVert[b]\rVert_{\w}^2
\ee
for all $[b]\in\mA/\mathcal{N}_{\w}.$
Therefore, $\pi_{w}(a)$ extends uniquely to a bounded operator on $\Hi_{\w}.$
It is immediate from (\ref{eq:GNSrepn}) that 
$\pi_{w}:\mA\to\mB(\Hi_{\w})$ is a
unital algebra homomorphism. It is a map of $C^*$-algebras because
\be
\begin{split}
\big\<[c],\pi_{\w}(a^*)[b]\big\>_{\w}
&=\big\<[c],[a^*b]\big\>_{\w}
=\w(c^*a^*b)
=\w\big((ac)^*b\big)\\
&=\big\<[ac],[b]\big\>_{\w}
=\big\<\pi_{\w}(a)[c],[b]\big\>_{\w}
\quad \forall \; a\in\mA,\;[b],[c]\in\mA/\mathcal{N}_{\w},
\end{split}
\ee
which shows that $\pi_{\w}(a^*)=\pi_{\w}(a)^*.$ 
Thus, associated to every state $\w:\mA\to\C,$
we have constructed a representation
$(\pi_{\w},\Hi_{\w})$ of $\mA.$ We denote
this assignment by
$\GNS_{\mA}:\St(\mA)\to\Rep(\mA),$
i.e. $\GNS_{\mA}(\w):=(\pi_{\w},\Hi_{\w}).$
It is automatically a functor because
$\St(\mA)$ has no non-trivial morphisms.
This construction is called the
\emph{\uline{GNS construction}} for $\mA.$
\econ

\begin{physics}
\label{phys:GNS}
As we discussed earlier, a state $\w:\mA\to\C$ is a list
of expectation values of all the observables
of interest described by $\mA.$
As a particular example, consider again
the case where $\mA=\mB(\Hi)$ for a Hilbert
space $\Hi$ with inner product
$\< \ \cdot \ , \ \cdot \ \>.$ Then,
there is actually a one-to-one
correspondence
between
states $\w:\mB(\Hi)\to\C$ satisfying a
certain condition%
\footnote{If
$\dim\Hi<\infty,$
no such additional condition is necessary.
However, in the case $\dim\Hi=\infty,$
one needs stronger, but reasonable, continuity
assumptions on the state.}
and 
density matrices, i.e. bounded
linear operators $\rho\in\mB(\Hi)$ 
that are self-adjoint and $\tr(\rho)=1$
(see Proposition 19.8 and Theorem 19.9 of \cite{HaQM}).
The correspondence is obtained
by the map that sends a density
matrix $\rho$ to the state $\w_{\rho}$
defined by $\w_{\rho}(a):=\tr(\rho a)$
for all $a\in\mA.$
Therefore, we will \emph{think}
of an abstract state $\w:\mA\to\C$
as being equivalent to a density matrix.%
\footnote{In some cases of interest,
such a density matrix need not exist. This
occurs for instance in the Unruh effect
whereupon restricting the algebra of
observables to a Rindler observer does
not lead to a density matrix, but rather an
abstract state satisfying the KMS condition
(see Section 5.1 of Wald \cite{Wa94}).}
This example will help us interpret
the GNS construction physically.
The meaning of the function
$(b,a)\mapsto\w(b^*a)$ for two
observables $a$ and $b$ in $\mA$ is
less mysterious if we 
focus on the case $b=a$ and think of $a$ and
$a^*$ as 
annihilation and creation operators, respectively.
Then $a^*a$ is the number operator and
$\w(a^*a)$ is the 
expectation value of the particle number
for the state $\w.$

The meaning of the null-space 
$\mathcal{N}_{\w}$ can be interpreted as the
set of observables that annihilate
the state $\w$ for all observable purposes.
If we go back to the case $\mA=\mB(\Hi)$
and the special case of 
$\rho=P_{\psi}$ (written as $|\psi\>\<\psi|$ in
Dirac notation),
the projection operator onto the subspace
spanned by a unit vector $\psi\in\Hi,$ then
for an observable $x\in\mA$ to be in $\mathcal{N}_{\w}$
would mean that 
$\tr(P_{\psi}x^*x)=\<x\psi,x\psi\>=0,$
which, since $\< \ \cdot \ , \ \cdot \ \>$ is an inner product,
would mean that $x\psi=0,$ i.e. $x$ annihilates
$\psi.$ Now, consider
two observables $a,b\in\mA$ such that
$b-a\in\mathcal{N}_{\w}.$ This means that
$(b-a)\psi=0,$ i.e. $b\psi=a\psi,$
which means that the observables
$b$ and $a$ cannot be distinguished by
the state $\<\psi,\;\cdot\;\psi\>$ associated to the vector $\psi.$ 
This argument extends to mixtures as well. 
The simple case of a density matrix of the form 
$\rho=\l P_{\psi}+(1-\l) P_{\phi},$ with $\psi,\phi\in\Hi$ unit vectors, 
$P_{\psi},P_{\phi}$ their associated projections, and $\l\in(0,1)$ 
illustrates the general case. Let $\w_{\rho}$ be the associated state
$\w_{\rho}=\tr(\rho\;\cdot\;).$ Then $x\in\mathcal{N}_{\w_{\rho}}$ implies
$\l\<x\psi,x\psi\>+(1-\l)\<x\phi,x\phi\>=0.$ Since all terms are non-negative,
this requires $\<x\psi,x\psi\>=0$ and $\<x\phi,x\phi\>=0$ individually. 
Hence, one concludes that $x\in\mathcal{N}_{\w_{\rho}}$ implies that $x$
annihilates \emph{all} the vectors comprising the mixture. 
Similarly, if $b-a\in\mathcal{N}_{\w_{\rho}},$ then one concludes
$a\psi=b\psi$ and $a\phi=b\phi$ so that $a$ and $b$ are
indistinguishable observables with respect to \emph{both} 
the vectors comprising the mixture.
Therefore, to summarize,
if we fix a state $\w$ on an algebra of observables
$\mA,$ it may be that with respect to that
particular state, there are some observables
that are indistinguishable in terms of their
expectation values. That is why we
consider the quotient $\mA/\mathcal{N}_{\w}$
where we have identified these 
equivalent observables.
Therefore, the GNS construction tells us that the 
associated Hilbert
space is just equivalence classes of observables
of $\mA$ distinguished by the state $\w.$
\end{physics}

\bcon
\label{con:GNSmorphisms}
Let $\mA'\xrightarrow{f}\mA$ be a morphism
of $C^*$-algebras and let $\w:\mA\to\C$ be a
state on $\mA.$ Then, as discussed
in Proposition \ref{prop:statefunctor},
$\w\circ f:\mA'\to\C$ is a state on $\mA'.$
By applying the previous construction,
we get two representations
$\pi_{\w\circ f}:\mA'\to\mB(\Hi_{\w\circ f})$
and
$\pi_{\w}:\mA\to\mB(\Hi_{\w})$
with $\pi_{\w\circ f}$ a representation of $\mA'$ and
$\pi_{\w}$ a representation of $\mA.$ 
There is a canonical map 
$L_{f}:\mA'/\mathcal{N}_{\w\circ f}\to\mA/\mathcal{N}_{\w}$
obtained from the diagram%
\footnote{The double arrow $\twoheadrightarrow$ signifies a surjection.}
\be
\label{eq:eq:GNSintertwinerdiagram}
\xy0;/r.25pc/:
(-20,7.5)*+{\mA'}="1";
(20,7.5)*+{\mA}="2";
(-20,-7.5)*+{\mA'/\mathcal{N}_{\w\circ f}}="3";
(20,-7.5)*+{\mA/\mathcal{N}_{\w}}="4";
{\ar@{->>}"1";"3"};
{\ar@{->>}"2";"4"};
{\ar"1";"2"^{f}};
{\ar@{-->}"3";"4"_{L_{f}}};
\endxy
\ee
given by
\be
\label{eq:GNSintertwiner}
L_{f}([a']):=[f(a')]
\ee
for all $[a']\in\mA'/\mathcal{N}_{\w\circ f}.$
This is well-defined because for any $x'\in\mathcal{N}_{\w\circ f},$%
\footnote{Diagrams such as 
(\ref{eq:GNSintertwinerwelldefined})
are read from top to bottom in either
clockwise or counterclockwise order to
replicate the argument in the order
in which it was originally conceived.}
\be
\label{eq:GNSintertwinerwelldefined}
\hspace{8mm}
\xy0;/r.30pc/:
(0,5)*+{(\w\circ f)\big(x'^{*}x'\big)}="1";
(-12.5,-5)*+{0}="4";
(12.5,-5)*+{\w\big(f(x')^*f(x')\big)}="2";
{\ar@{=}@/^1.0pc/"1";"2"^(0.70){\text{by Def'n \ref{defn:c*algebramorphism}}}};
{\ar@{=}@/^1.0pc/"4";"1"^(0.25){\text{since }
x'\in\mathcal{N}_{\w\circ f}}};
\endxy
\;\;,
\ee
i.e. $f(x')\in\mathcal{N}_{\w}.$
A similar calculation shows that 
\be
\label{eq:Lfisometry}
\left\lVert L_{f}\big([a']\big)\right\rVert_{\w}^2
=\left\lVert\big[f(a')\big]\right\rVert_{\w}^2
=\w\big(f(a')^*f(a')\big)
=(\w\circ f)\big(a'^{*}a'\big)
=\left\lVert[a']\right\rVert_{\w\circ f}^{2}
\ee
for all $[a']\in\mA'/\mathcal{N}_{\w\circ f}$ 
so that $L_{f}$ is an injective bounded linear map
and therefore extends uniquely to an injective bounded linear map
$L_{f}:\Hi_{\w\circ f}\to\Hi_{\w},$ which is denoted by the same letter. 
Furthermore, the map
$L_{f}$ is an intertwiner
$(\pi_{\w\circ f},\Hi_{\w\circ f})\to(\pi_{\w}\circ f,\Hi_{\w})$
of representations of $\mA',$
which means that
the diagram
\be
\xy0;/r.25pc/:
(-10,10)*+{\Hi_{\w\circ f}}="1";
(10,10)*+{\Hi_{\w}}="2";
(-10,-10)*+{\Hi_{\w\circ f}}="3";
(10,-10)*+{\Hi_{\w}}="4";
{\ar"1";"2"^{L_{f}}};
{\ar"3";"4"_{L_{f}}};
{\ar"1";"3"_{\pi_{\w\circ f}(a')}};
{\ar"2";"4"^{\pi_{\w}\big(f(a')\big)}};
\endxy
\ee
commutes for all $a'\in\mA'.$ This is true because
for any $[b']\in\mA'/\mathcal{N}_{\w\circ f},$
\be
\begin{split}
L_{f}\Big(\pi_{\w\circ f}(a')\big([b']\big)\Big)
&=L_{f}\big([a'b']\big)\qquad\text{ by (\ref{eq:GNSrepn}) for $\pi_{\w\circ f}$}\\
&=\big[f(a'b')\big]\qquad\text{ by (\ref{eq:GNSintertwiner})}\\
&=\big[f(a')f(b')\big]\qquad\text{ by Def'n \ref{defn:c*algebramorphism}}\\
&=\pi_{\w}\Big(f(a')\Big)\big([f(b')]\big)\qquad\text{ by (\ref{eq:GNSrepn}) for $\pi_{\w}$}\\
&=\pi_{\w}\big(f(a')\big)\Big(L_{f}\big([b']\big)\Big)\qquad\text{ by (\ref{eq:GNSintertwiner})}
\end{split}
\ee
for all $a'\in\mA'.$ 
By continuity, $L_{f}$ is an intertwiner on all of $\Hi_{\w\circ f}.$
The assignment sending a state $\w:\mA\to\C$
and a morphism $f:\mA'\to\mA$ of $C^*$-algebras
to the intertwiner 
$L_{f}:(\pi_{\w\circ f},\Hi_{\w\circ f})\to(\pi_{\w}\circ f,\Hi_{\w})$ 
of representations of $\mA'$
therefore defines a natural
transformation%
\footnote{In the present situation, the definition of the
natural transformation $\GNS_{f}$ reduces to an assignment
on objects of $\St(\mA)$ to morphisms of $\Rep(\mA')$ 
because $\St(\mA)$ is a discrete category.}
\be
\label{eq:GNSf}
\xy0;/r.25pc/:
(-15,10)*+{\St(\mA)}="1";
(15,10)*+{\Rep(\mA)}="2";
(-15,-10)*+{\St(\mA')}="3";
(15,-10)*+{\Rep(\mA')}="4";
{\ar"1";"2"^{\GNS_{\mA}}};
{\ar"3";"4"_{\GNS_{\mA'}}};
{\ar"1";"3"_{\St(f)}};
{\ar"2";"4"^{\Rep(f)}};
{\ar@{=>}"3";"2"|-{\GNS_{f}}};
\endxy
\ee
associated to every morphism
$f:\mA'\to\mA$ of $C^*$-algebras.
We denote the intertwiner $L_{f}$
by $\GNS_{f}(\w)$ to explicitly indicate what it depends on.
\econ

\begin{physics}
\label{phys:pullbackideal}
Let us go back to the case
$i:\mA_0\hookrightarrow\mA$ of restricting
ourselves to a subalgebra of observables and
let $\w$ be a state on $\mA.$ Let $\w_0:=\w\circ i$
be the state pulled back to $\mA_0.$
Since $\mA_0$ is a subalgebra of $\mA,$
there are
fewer experiments we can perform on the state $\w_{0}.$
Although
$\mathcal{N}_{\w_0}\subseteq\mathcal{N}_{\w},$
which was proven in (\ref{eq:GNSintertwinerwelldefined}), 
the fact that $L_{i}:\mA_{0}/\mathcal{N}_{\w_{0}}\to\mA/\mathcal{N}_{\w}$ 
is injective says that the equivalence classes of
distinguishable observables for the state $\w_{0}$ 
(cf. Physics \ref{phys:GNS}) are also
distinguishable by $\w,$ but not necessarily conversely. 
In this sense, there are fewer distinguishable observables 
for $\w_{0}$ than there are for $\w.$ 
This is consistent with the perspective that $\mA_{0}$ describes
macrostate observables while $\mA$ describes
microstate observables (cf. Physics \ref{phys:macrostates}). 
Since
the intertwiner provides a subspace 
$L_{i}(\Hi_{\w_0})$ of the
Hilbert space $\Hi_{\w},$  
the act of
restricting our view to a subalgebra
corresponds to restricting to a subspace of our
Hilbert space.%
\footnote{This phrasing is a bit misleading, however,
since \emph{every} $C^*$-algebra morphism
$f:\mA'\to\mA$ will lead to $L_{f}$ being injective
regardless of whether or not $f$ is injective
since our argument did not depend on this.
Nevertheless, for psychological reasons and simplicity
for interpretation, we will always use inclusions for
explaining the physics.}
\end{physics}

\bt
\label{thm:GNSfunctor}
The assignments%
\footnote{Given a 2-category (or a category)
$\mathcal{C},$ the objects, 1-morphisms, and
2-morphisms are denoted by $\mathcal{C}_0,$
$\mathcal{C}_1,$ and $\mathcal{C}_2,$ 
respectively.
}
\be
\begin{split}
\CAlg^{\op}_{0}&\xrightarrow{\GNS}\Cat_{1}\\
\mA&\mapsto
\Big(\St(\mA)\xrightarrow{\GNS_{\mA}}\Rep(\mA)\Big)
\end{split}
\ee
from Construction \ref{con:GNSobjects}
and
\be
\begin{split}
\CAlg^{\op}_{1}&\xrightarrow{\GNS}\Cat_{2}\\
\Big(\mA'\xrightarrow{f}\mA\Big)&\mapsto
\Big(\GNS_{\mA'}\circ\St(f)
\xRightarrow{\GNS_{f}}\Rep(f)\circ\GNS_{\mA}\Big)
\end{split}
\ee
from Construction \ref{con:GNSmorphisms}
define an oplax-natural transformation%
\footnote{We are viewing $\Cat$ as a 
strict 2-category whose 2-morphisms are natural
transformations. By also viewing $\CAlg^{\op}$ as
a 2-category (all of whose 2-morphisms are identities),
we can view $\St$ and $\Rep$ as (strict) 2-functors.
Because
$\GNS_{f}$ is not invertible, which is usually
required in the definition of a pseudo-natural
transformation,
we use the more general 
notion of oplax-natural transformation
described in
Definition 
\ref{defn:semipseudonaturaltransformation}
of the Appendix.
}
\be
\xy0;/r.25pc/:
(-20,0)*+{\CAlg^{\op}}="1";
(20,0)*+{\Cat}="2";
{\ar@/^1.5pc/"1";"2"^{\St}};
{\ar@/_1.5pc/"1";"2"_{\Rep}};
{\ar@{=>}(0,5);(0,-5)_{\GNS}};
\endxy
.
\ee
\et

\bprf
There are only two things to check because $\CAlg$ has no
nontrivial 2-morphisms (see
Definition \ref{defn:semipseudonaturaltransformation}).
First, the GNS construction
applied to the identity morphism
$\id_{\mA}$ for a $C^*$-algebra
$\mA$ gives $\GNS_{\id_{\mA}}$
which is precisely the identity
natural transformation
$\GNS_{\mA}\circ\St(\id_{\mA})
=\GNS_{\mA}\Rightarrow\GNS_{\mA}
=\Rep(\id_{\mA})\circ\GNS_{\mA}.$
Second,
associated to a pair of composable morphisms
\be
\mA''\xrightarrow{f'}\mA'\xrightarrow{f}\mA,
\ee
there are two diagrams one obtains. On the one
hand, applying the GNS construction
to the composition $f\circ f'$ gives $\GNS_{f\circ f'}.$
On the other hand, applying GNS to each $f'$
and $f$ and then composing gives another natural
transformation. These two results look like
\be
\label{eq:GNSpseudonaturalconditions}
\xy0;/r.25pc/:
(-15,10)*+{\St(\mA)}="1";
(15,10)*+{\Rep(\mA)}="2";
(-15,-10)*+{\St(\mA'')}="3";
(15,-10)*+{\Rep(\mA'')}="4";
{\ar"1";"2"^{\GNS_{\mA}}};
{\ar"3";"4"_{\GNS_{\mA''}}};
{\ar"1";"3"|-{\St(f\circ f')\quad}};
{\ar"2";"4"|-{\Rep(f\circ f')}};
{\ar@{=>}"3";"2"|-{\GNS_{f\circ f'}}};
\endxy
\quad\&\quad
\xy0;/r.25pc/:
(-15,20)*+{\St(\mA)}="1";
(15,20)*+{\Rep(\mA)}="2";
(-15,0)*+{\St(\mA')}="3";
(15,0)*+{\Rep(\mA')}="4";
(-15,-20)*+{\St(\mA'')}="5";
(15,-20)*+{\Rep(\mA'')}="6";
{\ar"1";"2"^{\GNS_{\mA}}};
{\ar"3";"4"_{\GNS_{\mA'}}};
{\ar"1";"3"|-{\St(f)}};
{\ar"2";"4"|-{\Rep(f)}};
{\ar@{=>}"3";"2"|-{\GNS_{f}}};
{\ar"5";"6"_{\GNS_{\mA''}}};
{\ar"3";"5"|-{\St(f')}};
{\ar"4";"6"|-{\Rep(f')}};
{\ar@{=>}"5";"4"|-{\GNS_{f'}}};
\endxy
,
\ee
respectively. The second condition that
$\GNS$ be an oplax-natural transformation
is that the compositions in these 
two diagrams are equal.
This follows from the commutativity of the 
individual squares and triangles in the diagram
\be
\xy0;/r.25pc/:
(-25,10)*+{\mA''}="1";
(0,10)*+{\mA'}="2";
(25,10)*+{\mA}="3";
(-25,-10)*+{\mA/\mathcal{N}_{\w\circ f\circ f'}}="4";
(0,-10)*+{\mA/\mathcal{N}_{\w\circ f}}="5";
(25,-10)*+{\mA/\mathcal{N}_{\w}}="6";
{\ar@{->>}"1";"4"};
{\ar@{->>}"2";"5"};
{\ar@{->>}"3";"6"};
{\ar"1";"2"_{f'}};
{\ar"2";"3"_{f}};
{\ar@/^1.5pc/"1";"3"^{f\circ f'}};
{\ar@{-->}"4";"5"^{L_{f'}}};
{\ar@{-->}"5";"6"^{L_{f}}};
{\ar@/_1.5pc/@{-->}"4";"6"_{L_{f\circ f'}}};
\endxy
\ee
for any state $\w$ on $\mA.$ 
By continuity, this equality extends to the completions.
\eprf

\begin{physics}
Oplax-naturality means the following
if we restrict our attention
to a subalgebra and then restrict to yet another
subalgebra, as in 
\be
\mA_1 \xhookrightarrow{j} \mA_0 \xhookrightarrow{i} \mA.
\ee
Equality of the two diagrams in
(\ref{eq:GNSpseudonaturalconditions})
means that constructing the
physical subspace $\Hi_{\w\circ i\circ j}$
of $\Hi_{\w}$
of quantum configurations
for the state $\w$ with respect to the subalgebra
$\mA_{1}$ is the same subspace obtained
from first restricting to $\mA_{0}$ and then
to $\mA_{1},$ i.e.
\be
\xy0;/r.25pc/:
(-10,-7.5)*+{\Hi_{\w\circ i\circ j}}="1";
(10,-7.5)*+{\Hi_{\w}}="3";
(0,7.5)*+{\Hi_{\w\circ i}}="2";
{\ar"1";"3"_{L_{i\circ j}}};
{\ar"1";"2"^{L_{j}}};
{\ar"2";"3"^{L_{i}}};
\endxy
\ee
commutes, where we have used the notation from
Construction \ref{con:GNSmorphisms}.
\end{physics}

\br
$\GNS$ being an oplax-natural transformation 
provides the correct categorical structured needed to reflect the 
functoriality of the GNS construction as can be seen
by the equality of the diagrams in 
(\ref{eq:GNSpseudonaturalconditions}). 
In current terminology \cite{Br93}, Theorem \ref{thm:GNSfunctor} shows
that the GNS construction is not only
a functor for a fixed $C^*$-algebra $\mA,$ but it is also a morphism
of pre-stacks over the category of all $C^*$-algebras. Note that it is \emph{not}
a morphism of pre-sheaves of categories because
the outer diagram in (\ref{eq:GNSf}) does
\emph{not} commute (a condition that
is required to have a morphism of
pre-sheaves). Instead, a natural transformation
(which is a 2-morphism in $\Cat$) 
is needed to compensate for the lack of commutativity,
and this is why 2-categories play a crucial role in the GNS construction.
\er

\section{Some comments on the category of states}
\label{sec:comments}

One would like to think of $\St(\mA)$ as a category
of states with non-trivial morphisms. Namely,
a morphism from $\w:\mA\to\C$ to $\mu:\mA\to\C$
consists of a $C^*$-algebra morphism
$\f:\mA\to\mA$ such that the diagram
\be
\xy0;/r.25pc/:
(-10,7.5)*+{\mA}="1";
(10,7.5)*+{\mA}="2";
(0,-7.5)*+{\C}="3";
{\ar"1";"3"_{\w}};
{\ar"1";"2"^{\f}};
{\ar"2";"3"^{\mu}};
\endxy
\ee
commutes, i.e. $\mu\circ\f=\w.$
Let us call this closely related category
$\states\;(\mA).$
While one can define a functor
$\states\;(\mA)\to\Rep(\mA)$
as a special case
of $\GNS_{\mA}$ on objects
and $\GNS_{\f}$ on morphisms,
this is too restrictive and not
what we want in general because
mappings of \emph{different} algebras show up in many applications. 
Recall from Physics \ref{phys:macrostates}, that
a $C^*$-algebra map
$\mA_{0}\to\mA$ is
supposed to be thought of as
using macrostate observables
described by $\mA_0$
instead of microstate observables
described by $\mA.$
To incorporate this, we would therefore
still want to think of the different categories
of states as a pre-sheaf of categories on
the category of $C^*$-algebras, i.e. a functor
$\states:\CAlg^{\op}\to\Cat.$ 
For a morphism $f:\mA'\to\mA$ this should
get mapped to a functor
$\states\;(f):\states\;(\mA)\to\states\;(\mA').$
How should this functor be defined? This
agrees with $\St(f)$ at the level of objects.
However, for a morphism $\f:\w\to\mu$
of states in $\mA,$ all we have is the
collection of morphisms
\be
\xy0;/r.25pc/:
(-20,15)*+{\mA'}="a";
(20,15)*+{\mA'}="b";
(-10,0)*+{\mA}="1";
(10,0)*+{\mA}="2";
(0,-15)*+{\C}="3";
{\ar"a";"1"_{f}};
{\ar"b";"2"^{f}};
{\ar"1";"3"_{\w}};
{\ar"1";"2"^{\f}};
{\ar"2";"3"^{\mu}};
\endxy
.
\ee
From these data, we are supposed to produce
a map of $C^*$-algebras $\f':\mA'\to\mA'$ such that
the diagram
\be
\xy0;/r.25pc/:
(-10,7.5)*+{\mA'}="1";
(10,7.5)*+{\mA'}="2";
(0,-7.5)*+{\C}="3";
{\ar"1";"3"_{\w\circ f}};
{\ar"1";"2"^{\f'}};
{\ar"2";"3"^{\mu\circ f}};
\endxy
\ee
commutes. One can show that the only such
maps $f:\mA'\to\mA$ of $C^*$-algebras
for which we can do this in a functorial
manner are $C^*$-algebra
\emph{isomorphisms}. Since we specifically
do not want this for physical reasons, we
use the discrete category $\St(\mA)$
instead of the more reasonable, yet naive,
category $\states\;(\mA).$

\section{A right adjoint to the GNS construction}
\label{sec:adjoint}

Besides producing a representation 
$(\pi_{\w},\Hi_{\w})$ of $\mA$
given a state $\w$ on $\mA,$ the GNS construction
also produces a cyclic vector in $\Hi_{\w}.$ This fact
will let us construct a sort of 
inverse to the GNS construction
provided that we include this extra datum in the
definition of the oplax-natural transformation
$\GNS.$ 
\bd
\label{defn:cyclic}
A \emph{\uline{cyclic vector}} $\W$ 
for a representation $\pi$
of a $C^*$-algebra $\mA$ on a Hilbert space $\Hi$
is a normalized (i.e. unit) vector $\W\in\Hi$ such that
\be
\{ \pi(a)\W \; : \; a\in\mA \}
\ee
is a dense subset in $\Hi$ (with respect
to the norm induced by the inner product
on $\Hi$). 
A representation $(\pi,\Hi)$ of $\mA$ 
together with a cyclic vector $\W$ is called a
\emph{\uline{cyclic representation}}
and is written as a triple $(\pi,\Hi,\W).$
A representation $(\pi,\Hi)$ of $\mA$
together with a normalized vector (not necessarily cyclic)
is called a \emph{\uline{pointed representation}}.
\ed

The reason for demanding normalized vectors in the above definition
is so that they produce states, as will be explained shortly. 

\begin{physics}
\label{phys:cyclicgroundstate}
When $\mA$ is the algebra of observables
for a quantum field theory (in a particular phase), the vacuum vector
is typically a cyclic vector---any particle content
state is obtained by creation operators on the ground state.
When a representation is irreducible, every
non-zero vector is cyclic---by using annihilation operators, 
one can get to the ground state. 
One should take this comment with a grain of salt due
to domain and distribution issues (see Folland \cite{Fo08}).
A finite-dimensional and completely rigorous example occurs
in the theory of spin by use of ladder operators.
\end{physics}

\bd
\label{defn:pointedRepobjects}
Let $\pRep(\mA)$ be the category
of pointed representations of $\mA.$
Namely, an object of 
$\pRep(\mA)$ is a pointed representation $(\pi,\Hi,\W)$ of
$\mA$ and a morphism from $(\pi,\Hi,\W)$ to $(\pi',\Hi',\W')$
is an intertwiner $L:\Hi\to\Hi'$ of
representations of $\mA$ such that 
\be
\label{eq:morphismspointedreps}
L(\W)=\W'
\aand
L^*L=\id_{\Hi}.
\ee 
Let $\cRep(\mA)$ 
be the sub-category of $\pRep(\mA)$
of cyclic representations of $\mA.$ 
\ed

\bprf
Some things must be checked so that the above
definition is in fact valid. For instance, let
\be
(\pi,\Hi,\W)\xrightarrow{L}(\pi',\Hi',\W')
\xrightarrow{L'}(\pi'',\Hi'',\W'')
\ee
be a pair of composable morphisms. Then the
composition $L'L$ satisfies
\be
(L'L)^*(L'L)=L^*L'^*L'L=L^*L=\id_{\Hi}.
\ee
Associativity follows from associativity of 
composition of functions. The other
axioms of a category all hold.
\eprf

\br
$\cRep(\mA)$ 
is a full subcategory of $\pRep(\mA)$
because a vector being cyclic is a property
and not additional structure.
\er

The following lemma will be useful for proving that certain
linear maps are isometries. 

\blem
\label{lem:isometry}
Let $\Hi$ and $\Hi'$ be Hilbert spaces and let $L:\Hi\to\Hi'$
be a bounded linear map. The following conditions on $L$ are equivalent. 
\begin{enumerate}[(a)]
\item $L^*L=\id_{\Hi}.$
\item $\lVert L\psi\rVert'=\lVert\psi\rVert$ for all $\psi\in\Hi.$
\item $\<L\psi,L\phi\>'=\<\psi,\phi\>$ for all $\psi,\phi\in\Hi.$
\end{enumerate}
In this notation, a prime superscript has been used to distinguish
the norm and inner product on $\Hi'$ from that of $\Hi.$ 
\elem

\br
\label{rmk:dense}
The condition $L^*L=\id_{\Hi}$ implies that $L$ is injective.
We do \emph{not} require $L$ to be unitary,
which would impose $LL^*=\id_{\Hi'}$ as well.
However, note that if $L:(\pi,\Hi,\W)\to(\pi',\Hi',\W')$
is a morphism of \emph{cyclic} representations, then
$L$ sends a dense subset of $\Hi$ to a 
dense subset of $\Hi'$ because $L(\W)=\W'.$
Therefore, in this case, $L$ is unitary.
Also note that the number of morphisms from a cyclic representation
to a pointed representation is quite small: there is either
one or none at all. 
\er

\bcon
\label{con:restrictionfunctor}
Let $(\pi,\Hi,\W)$ be a pointed representation
of a $C^*$-algebra $\mA.$ The vector $\W$
defines a state $\w_{\W}$ on $\mB(\Hi)$ by 
the formula
\be
\label{eq:omegaOmega}
\mB(\Hi)\ni B\mapsto\w_{\W}(B):=\<\W,B\W\>.
\ee
We often refer to $\w_{\W}$ as a vector state. 
Pulling this state back to $\mA$ along $\pi$ defines a
state $\w_{\W}\circ\pi:\mA\to\C$ on $\mA.$ 
$\w_{\W}\circ\pi$ is sometimes denoted by
$\rest_{\mA}\big((\pi,\Hi,\W)\big)$ for ``restriction.''
\econ

\br
If the vector $\W$ were not required to be normalized, but merely
nonzero, then
one could still define a state by the assignment
\be
\mB(\Hi)\ni B\mapsto\frac{\<\W,B\W\>}{\<\W,\W\>}.
\ee
Nevertheless, such an assignment would spoil other
desirable properties that will be discussed shortly
(see Remark \ref{rmk:normalizationmodification}). 
\er

\blem
\label{lem:restmorphisms}
Let $L:(\pi,\Hi,\W)\to(\pi',\Hi',\W')$
be a morphism of pointed representations
of $\mA.$
Then,
\be
\w_{\W'}\circ\pi'=\w_{\W}\circ\pi,
\ee
i.e. the two states 
$\rest_{\mA}\big((\pi,\Hi,\W)\big)$
and
$\rest_{\mA}\big((\pi',\Hi',\W')\big)$
are equal.
\elem

\bprf
For any $a\in\mA,$
\be
\begin{split}
\left\<\W',\pi'(a)\W'\right\>
&=\<L(\W),\pi'(a)L(\W)\>\qquad\text{ by (\ref{defn:pointedRepobjects})}\\
&=\<L(\W),L\pi(a)\W\>\qquad\text{ by (\ref{eq:intertwiner})}\\
&=\<\W,L^{*}L\pi(a)\W\>\qquad\text{ by Def'n of $L^*$}\\
&=\<\W,\pi(a)\W\>\qquad\text{ by (\ref{defn:pointedRepobjects})}.
\end{split}
\ee
\eprf

\begin{physics}
Imagine a context in which
we begin with a representation of $\mA$ on a Hilbert space $\Hi$
and a vacuum vector $\W\in\Hi.$ Given
a subalgebra $\mA$ of $\mB(\Hi),$ the
construction $\rest_{\mA}$ restricts the
vacuum state $\w_{\W}:=\<\W,\;\cdot\;\W\>$ on $\mB(\Hi)$  
to a state on this subalgebra.
This is useful if we can only make
measurements of certain observables.
For instance, a Rindler observer has a restricted
algebra of observables so that restricting
a Minkowski vacuum state to their
algebra results in a thermal state, a
phenomenon known as the Unruh effect \cite{Wa94}.
If we change our representation
in such a way
that the two are still related by an intertwiner
satisfying (\ref{eq:morphismspointedreps}),
then we get the \emph{same} state. 
Note that it is not required that $L$ be unitary---an isometry
preserving the unit vector suffices. 
$\rest_{\mA}$ is also a natural construction from
the physics perspective since every 
unit vector in $\Hi$ gives a state on
any $C^*$-subalgebra of $\mB(\Hi).$
What is not obvious is that there is 
a canonical way to go back---the purpose of this
section is to make this statement precise
and prove that the GNS construction
achieves this. 
\end{physics}

\bn
Let $\mA$ be a $C^*$-algebra. 
The assignment 
\be
\begin{split}
\pRep(\mA)_{0}\ni(\pi,\Hi,\W)
&\mapsto\w_{\W}\circ\pi\in\St(\mA)_{0}\\
\pRep(\mA)_{1}\ni 
\Big((\pi,\Hi,\W)\xrightarrow{L}(\pi',\Hi',\W')\Big)
&\mapsto\id_{\w_{\W}\circ\pi}\in\St(\mA)_{1}
\end{split}
\ee
from Construction \ref{con:restrictionfunctor}
defines a functor
$\rest_{\mA}:\pRep(\mA)
\to\St(\mA).$
\en

\bprf
This follows directly
from Construction \ref{con:restrictionfunctor}
and Lemma \ref{lem:restmorphisms}.
\eprf

\bcon
\label{con:pointedRepmorphisms}
Let $\mA'\xrightarrow{f}\mA$ be a morphism
of $C^*$-algebras. The induced functor
$\Rep(f):\Rep(\mA)\to\Rep(\mA')$ extends
to a functor 
$\pRep(f):
\pRep(\mA)
\to\pRep(\mA')$ as follows.
Let $(\pi,\Hi,\W)$ be a
pointed representation of $\mA.$ Then
this gets sent to $(\pi\circ f,\Hi,\W).$
Note that even if $(\pi,\Hi,\W)$ is a cyclic
representation, 
$(\pi\circ f,\Hi,\W)$ is \emph{not} necessarily a cyclic
representation of $\mA'$ since
\be
\left\{\pi\big(f(a')\big)\W \; : \; a'\in\mA'\right\}
\ee
is not necessarily dense in $\Hi.$
Nevertheless, $(\pi\circ f,\Hi,\W)$ \emph{is}
a pointed representation.
A morphism of pointed representations
of $\mA$ gets sent to a morphism of
pointed representations of $\mA'$
under the functor $\pRep(f)$
using the same intertwiner. In fact,
the diagram
\be
\label{eq:restnaturaltransformation}
\xy0;/r.25pc/:
(-15,10)*+{\pRep(\mA)}="1";
(-15,-10)*+{\pRep(\mA')}="2";
(15,10)*+{\St(\mA)}="3";
(15,-10)*+{\St(\mA')}="4";
{\ar"1";"3"^{\rest_{\mA}}};
{\ar"2";"4"_{\rest_{\mA'}}};
{\ar"1";"2"_{\pRep(f)}};
{\ar"3";"4"^{\St(f)}};
\endxy
\ee
commutes.
\econ

This proves the following fact.

\bn
\label{prop:rest}
$\rest,$ as defined in Construction 
\ref{con:restrictionfunctor},
is a natural transformation%
\footnote{This is special case of a
pseudo-natural transformation
since $\rest_{f}=\id$ in
(\ref{eq:restnaturaltransformation}).
}
\be
\xy0;/r.25pc/:
(-20,0)*+{\CAlg^{\op}}="1";
(20,0)*+{\Cat}="2";
{\ar@/^1.5pc/"1";"2"^{\St}};
{\ar@/_1.5pc/"1";"2"_{\pRep}};
{\ar@{=>}(0,-5);(0,5)_{\rest}};
\endxy
.
\ee
\en

\begin{physics}
Following the earlier examples of 
a subalgebra $\mA_{0}\hookrightarrow\mA,$ 
one interpretation of the functor 
$\pRep(\mA_{0}\hookrightarrow\mA):\pRep(\mA)\to\pRep(\mA_0)$ can
be given as follows. 
Let $(\pi,\Hi,\W)$ be a cyclic representation of $\mA$ 
with $\W$ viewed as a vacuum vector for some quantum field theory.
View $\mA_{0}$ as a subalgebra of $\mA$ corresponding
to low energy observables. Low energy observables
may be unable to produce states of the parent theory whose rest mass
is beyond the energy scale of the low energy observer. 
This corresponds to the fact that the vector $\W$ might no 
longer be cyclic with respect to the pullback representation
to the algebra $\mA_{0}$ (cf. Physics \ref{phys:cyclicgroundstate}). 

Commutativity of the diagram (\ref{eq:restnaturaltransformation}),
and hence naturality of $\rest$, says that the state the low 
energy observer sees is independent of the order in which they
disregard certain information. The observer can either 
forget the representation first and then focus on the 
low energy observables or first focus on the induced low energy
representation and then forget the representation. 
The induced states are the same. 
\end{physics}

We will now modify the GNS construction to
include the construction of a cyclic vector.
Due to the similarity of this construction
and that of Constructions \ref{con:GNSobjects}
and \ref{con:GNSmorphisms},
we will skip many details and only focus on
the new ones.

\bcon
\label{con:GNScyclicpseudo}
For every $C^*$-algebra $\mA,$ define a functor 
\be
\pGNS_{\mA}:\St(\mA)\to\pRep(\mA)
\ee
by the following assignment. To a state
$\w:\mA\to\C,$ assign 
the cyclic representation%
\footnote{$[1_{\mA}]$ is a cyclic vector
because $\{\pi_{\w}(a)[1_{\mA}]=[a] \;:\; a\in\mA\}
=:\mA/\mathcal{N}_{\w}$
is dense in $\Hi_{\w}$ by definition.
}
$\pGNS_{\mA}(\w):=(\pi_{\w},\Hi_{\w},[1_{\mA}]).$ 
Because $\St(\mA)$ has no non-trivial morphisms,
this defines a functor.
Furthermore, the image of this functor actually lands
in the subcategory 
$\cRep(\mA)$ \cite{Se47}.
To every morphism $\mA'\xrightarrow{f}\mA$ 
of unital $C^*$-algebras, 
define a natural transformation
\be
\label{eq:GNScyclicf}
\xy0;/r.25pc/:
(-15,10)*+{\St(\mA)}="1";
(15,10)*+{\pRep(\mA)}="2";
(-15,-10)*+{\St(\mA')}="3";
(15,-10)*+{\pRep(\mA')}="4";
{\ar"1";"2"^{\pGNS_{\mA}}};
{\ar"3";"4"_{\pGNS_{\mA'}}};
{\ar"1";"3"_{\St(f)}};
{\ar"2";"4"^{\pRep(f)}};
{\ar@{=>}"3";"2"|-{\pGNS_{f}}};
\endxy
\ee
as follows. To every state $\w:\mA\to\C$ on
$\mA$ define the morphism
\be
\label{eq:GNSmonotonic}
\big(\pi_{\w\circ f},\Hi_{\w\circ f},[1_{\mA'}]\big)
\xrightarrow{\pGNS_{f}(\w)}
\big(\pi_{\w}\circ f,\Hi_{\w},[1_{\mA}]\big)
\ee
of pointed representations
to be exactly the same as $L_{f}$ 
in (\ref{eq:GNSintertwiner})
and simply note that a property of this linear
map is that
\be
L_{f}([1_{\mA'}])=[f(1_{\mA'})]=[1_{\mA}]
\ee
since $f$ is a morphism of \emph{unital} $C^*$-algebras.
$L_{f}^* L_{f}=\id_{\Hi_{\w\circ f}}$ 
follows from the calculation (\ref{eq:Lfisometry}) 
and Lemma \ref{lem:isometry}. 
\econ

\br
Note that although  
$\big(\pi_{\w},\Hi_{\w},[1_{\mA}]\big)$ and
$\big(\pi_{\w\circ f},\Hi_{\w\circ f},[1_{\mA'}]\big)$
are cyclic representations of $\mA$ and $\mA',$
respectively,
the pointed representation 
$\big(\pi_{\w}\circ f,\Hi_{\w},[1_{\mA}]\big)$
of $\mA'$ 
obtained by pullback along $f$ is \emph{not}
necessarily cyclic. This is why the target of
the GNS functor was chosen to be
the category of pointed representations instead
of cyclic representations. This is analogous 
to the fact that the restriction of an irreducible
representation to a subalgebra need not be
irreducible.
\er

\bt
\label{thm:pointedGNS}
The assignments
\be
\begin{split}
\CAlg^{\op}_{0}&\xrightarrow{\pGNS}\Cat_{1}\\
\mA&\mapsto
\Big(\St(\mA)\xrightarrow{\pGNS_{\mA}}
\pRep(\mA)\Big)
\end{split}
\ee
and
\be
\begin{split}
\CAlg^{\op}_{1}&\xrightarrow{\pGNS}\Cat_{2}\\
\Big(\mA'\xrightarrow{f}\mA\Big)&\mapsto
\Big(\pGNS_{\mA'}\circ\St(f)
\xRightarrow{\pGNS_{f}}\pRep(f)\circ
\pGNS_{\mA}\Big)
\end{split}
\ee
from Construction \ref{con:GNScyclicpseudo}
define an oplax-natural transformation
\be
\xy0;/r.25pc/:
(-20,0)*+{\CAlg^{\op}}="1";
(20,0)*+{\Cat}="2";
{\ar@/^1.5pc/"1";"2"^{\St}};
{\ar@/_1.5pc/"1";"2"_{\pRep}};
{\ar@{=>}(0,5);(0,-5)_{\pGNS}};
\endxy
.
\ee
\et
\bprf
The proof is not much different than what 
it was for $\GNS$ in Theorem \ref{thm:GNSfunctor}. 
\eprf

\begin{physics}
The weak form of naturality for $\pGNS$ guarantees
that the way in which the Hilbert spaces and their cyclic vectors
fit into the larger space is independent of how the restrictions to 
successive subalgebras is grouped. For instance, one could
imagine a sequence of course-grained subalgebras. This naturality
on one pair of subalgebras guarantees consistency for all such
tuples of subalgebras when applying the GNS construction. 
\end{physics}

There is one last construction we must
confront. This involves relating
the composition of 
oplax-natural transformations
$\rest$ and $\pGNS$ 
with the identity natural transformation.

\blem
The vertical composition
\be
\xy0;/r.25pc/:
(-20,0)*+{\CAlg^{\op}}="1";
(20,0)*+{\Cat}="2";
{\ar@/^2.5pc/"1";"2"^{\St}};
{\ar"1";"2"|-(0.4){\pRep}};
{\ar@/_2.5pc/"1";"2"_{\St}};
{\ar@{=>}(0,9);(0,2)_{\pGNS}};
{\ar@{=>}(0,-2);(0,-9)_{\rest}};
\endxy
\ee
of oplax-natural transformations is equal to the identity natural transformation.
\elem

\bprf
Let $\mA$ be a $C^*$-algebra.
The composition acting on a state $\w:\mA\to\C$
gives
\be
\begin{split}
&
\xy0;/r.25pc/:
(-30,0)*+{\St(\mA)}="1";
(0,0)*+{\pRep(\mA)}="2";
(30,0)*+{\St(\mA)}="3";
{\ar"1";"2"^(0.525){\pGNS_{\mA}}}
{\ar"2";"3"^(0.45){\rest_{\mA}}}
\endxy
\\
&\qquad
\xy0;/r.25pc/:
(-25,0)*+{\w}="1";
(5,0)*+{\big(\pi_{\w},\Hi_{\w},[1_{\mA}]\big)}="2";
(40,0)*+{\big\<[1_{\mA}],\pi_{\w}(\ \cdot \ ) 
[1_{\mA}]\big\>_{\w}}="3";
{\ar@{|->}"1";"2"}
{\ar@{|->}"2";"3"}
\endxy
.
\end{split}
\ee
$\big\<[1_{\mA}],\pi_{\w}(\ \cdot \ )[1_{\mA}]\big\>_{\w}$ 
agrees with $\w$ as a state on $\mA$ because
\be
\label{eq:statestoreptostates}
\big\<[1_{\mA}],\pi_{\w}(a) 
[1_{\mA}]\big\>_{\w}
=\big\<[1_{\mA}],[a]\big\>_{\w}
=\w(1_{\mA}^*a)
=\w(a)
\quad \forall \; a\in\mA.
\ee
There are no non-trivial morphisms 
in $\St(\mA)$ so the composition is the identity
functor.
To every morphism $f:\mA'\to\mA$ of $C^*$-algebras,
the composition of natural transformations
\be
\xy0;/r.25pc/:
(-30,10)*+{\St(\mA)}="1";
(0,10)*+{\pRep(\mA)}="2";
(-30,-10)*+{\St(\mA')}="3";
(0,-10)*+{\pRep(\mA')}="4";
(30,10)*+{\St(\mA)}="5";
(30,-10)*+{\St(\mA')}="6";
{\ar"1";"2"^{\pGNS_{\mA}}};
{\ar"3";"4"_{\pGNS_{\mA'}}};
{\ar"1";"3"_{\St(f)}};
{\ar"2";"4"|-{\pRep(f)}};
{\ar"2";"5"^{\rest_{\mA}}};
{\ar"4";"6"_{\rest_{\mA'}}};
{\ar"5";"6"^{\St(f)}};
{\ar@{=>}"3";"2"|-{\pGNS_{f}}};
{\ar@{=}"4";"5"|-{\id=\rest_{f}}};
\endxy
\ee
must equal the identity natural transformation.
This follows immediately from the fact that the compositions
on the top and bottom of this diagram are 
identities by (\ref{eq:statestoreptostates})
and since $\St(\mA')$ has no non-trivial morphisms.
\eprf

\begin{physics}
The interpretation of this is immediate when viewing $\pGNS$ and
$\rest$ as processes/constructions. If you start with a state,
construct the GNS representation, and then forget the representation,
you get back your original state. In other words, there is no loss
of information. 
\end{physics}

However, the composition in the order
\be
\xy0;/r.25pc/:
(-20,0)*+{\CAlg^{\op}}="1";
(20,0)*+{\Cat}="2";
{\ar@/^2.5pc/"1";"2"^{\pRep}};
{\ar"1";"2"|-(0.4){\St}};
{\ar@/_2.5pc/"1";"2"_{\pRep}};
{\ar@{=>}(0,9);(0,2)_{\rest}};
{\ar@{=>}(0,-2);(0,-9)_{\pGNS}};
\endxy
\ee
is certainly not the identity. In the following, 
we construct the required modification
(see Definition \ref{defn:GNSmodification}
in the Appendix).

\bcon
\label{con:modification}
Let $\mA$ be a unital $C^*$-algebra
and consider the diagram
\be
\xy0;/r.25pc/:
(-15,-7.5)*+{\pRep(\mA)}="1";
(0,7.5)*+{\St(\mA)}="2";
(15,-7.5)*+{\pRep(\mA)}="3";
{\ar"1";"2"^(0.475){\rest_{\mA}}};
{\ar"2";"3"^(0.55){\pGNS_{\mA}}};
{\ar"1";"3"_{\id_{\pRep(\mA)}}};
\endxy
\ee
of functors. 
Recalling the notation from Constructions
\ref{con:restrictionfunctor} and \ref{con:GNSobjects},
observe what happens to a pointed
representation
$(\pi,\Hi,\W)$ of $\mA$ along the top two functors
\be
\label{eq:restthenGNS}
\xy0;/r.25pc/:
(-40,0)*+{(\pi,\Hi,\W)}="1";
(0,0)*+{\big\<\W,\pi(\ \cdot \ )\W\big\>=:\w}="2";
(45,0)*+{\big(\pi_{\w},\Hi_{\w},[1_{\mA}]\big)}="3";
{\ar@{|->}"1";"2"^(0.425){\rest_{\mA}}};
{\ar@{|->}"2";"3"^(0.525){\pGNS_{\mA}}};
\endxy
\ee
Therefore, we have two pointed representations of $\mA$ 
whose associated states agree, i.e.  
\be
\label{eq:Segalcondition}
\<\W,\pi(a)\W\>
=\big\<[1_{\mA}],\pi_{\w}(a)[1_{\mA}]\big\>_{\w}
\quad \mbox{ for all } a\in\mA.
\ee
\emph{If} $(\pi,\Hi,\W)$ were \emph{also} a cyclic
representation, then 
it was already known by Segal
that any other cyclic representation
restricting to the same state is unitarily
equivalent to it \cite{Se47}.
We slightly modify Segal's proof for our construction. 
Define the linear map
\be
\label{eq:Segalmap}
\begin{split}
\mA/\mathcal{N}_{\w}
&\xrightarrow{m_{\mA}\big((\pi,\Hi,\W)\big)}\Hi\\
[a]&\xmapsto{\qquad\qquad\ \ }\pi(a)\W
\end{split}
\;\;\;.
\ee
To see that this is well-defined, let $x\in\mathcal{N}_{\w}$ so that
\be
\xy0;/r.20pc/:
(0,12.5)*+{\w(x^*x)}="1";
(15,0)*+{\big\<\W,\pi(x^*x)\W\big\>}="2";
(0,-12.5)*+{\big\<\pi(x)\W,\pi(x)\W\big\>}="3";
(-15,)*+{0}="4";
{\ar@{=}@/^0.8pc/"1";"2"^(0.7){\text{by (\ref{eq:restthenGNS})}}};
{\ar@{=}@/^0.8pc/"2";"3"};
{\ar@{=}@/^0.8pc/"4";"1"};
\endxy
\;\;\;.
\ee
Since $\< \ \cdot \ , \ \cdot \ \>$ is an inner product,
this holds if and only if $\pi(x)\W=0,$ proving well-definedness. 
Because%
\footnote{We occasionally write
$m_{\mA}([a])$ instead of
$m_{\mA}\big((\pi,\Hi,\W)\big)([a])$
because the notation would be too difficult
to read otherwise. Since our representation
$(\pi,\Hi,\W)$
is fixed for now, this should cause no confusion.
}
\be
\label{eq:Segalisometry}
\left\lVert m_{\mA}\big([a]\big)\right\rVert^2
=\big\<m_{\mA}\big([a]\big),m_{\mA}\big([a]\big)\big\>
=\big\<\pi(a)\W,\pi(a)\W\big\>
=\w(a^*a)
=\lVert[a]\rVert_{\w}^{2}
\ee
for all $[a]\in\mA/\mathcal{N}_{\w},$
$m_{\mA}$ is bounded and extends uniquely to 
a bounded linear map 
$m_{\mA}\big((\pi,\Hi,\W)\big):\Hi_{\w}\to\Hi.$
By Lemma \ref{lem:isometry} and (\ref{eq:Segalisometry}), 
this extension is also an isometry.

We now show that 
$m_{\mA}\big((\pi,\Hi,\W)\big):\big(\pi_{\w},\Hi_{\w},[1_{\mA}]\big)\to(\pi,\Hi,\W)$ 
is an intertwiner, which means that the diagram
\be
\label{eq:segalintertwiner}
\xy0;/r.25pc/:
(-15,7.5)*+{\Hi_{\w}}="1";
(15,7.5)*+{\Hi}="2";
(-15,-7.5)*+{\Hi_{\w}}="3";
(15,-7.5)*+{\Hi}="4";
{\ar"1";"2"^{m_{\mA}\big((\pi,\Hi,\W)\big)}};
{\ar"3";"4"_{m_{\mA}\big((\pi,\Hi,\W)\big)}};
{\ar"1";"3"_{\pi_{\w}(a)}};
{\ar"2";"4"^{\pi(a)}};
\endxy
\ee
commutes for all $a\in\mA.$ Following
the image of an element $[b]\in\mA/\mathcal{N}_{\w}$
along both of these morphisms gives
\be
\label{eq:Segalsproof}
\xy0;/r.20pc/:
(11.7557,16.1803)*+{\qquad\quad
\pi(a)m_{\mA}\big([b]\big)}="1";
(19.0211,-6.1803)*+{\pi(a)\pi(b)\W}="2";
(0,-20)*+{\pi(ab)\W}="3";
(-19.0211,-6.1803)*+{m_{\mA}\big([ab]\big)}="4";
(-11.7557,16.1803)*+{m_{\mA}\big(\pi_{\w}(a)[b]\big)
\quad\qquad}="5";
{\ar@{=}@/^0.7pc/"1";"2"^(0.55){\text{ by (\ref{eq:Segalmap})}}};
{\ar@{=}@/^0.7pc/"2";"3"};
{\ar@{=}@/^0.7pc/"3";"4"^(0.65){\text{ by (\ref{eq:Segalmap})}}};
{\ar@{=}@/^0.7pc/"4";"5"^(0.45){\text{ by (\ref{eq:GNSrepn})}}};
\endxy
\ee
proving that the diagram (\ref{eq:segalintertwiner}) indeed commutes 
(upon extension to the completion).
Hence, $m_{\mA}\big((\pi,\Hi,\W)\big):\big(\pi_{\w},\Hi_{\w},[1_{\mA}]\big)\to(\pi,\Hi,\W)$
is a morphism in $\pRep(\mA).$

Note that if $(\pi,\Hi,\W)$ is cyclic, then 
Remark \ref{rmk:dense} shows that $m_{\mA}\big((\pi,\Hi,\W)\big)$ 
is a unitary equivalence. 
\econ

\br
\label{rmk:normalizationmodification}
If, in the definition of a pointed representation, we use 
arbitrary vectors instead of normalized ones and we define
$\rest_{\mA}(\pi,\Hi,\W)$ to be the state 
\be
\frac{\<\W,\pi(\ \cdot \ )\W\>}{\<\W,\W\>},
\ee
then the map (\ref{eq:Segalmap}) cannot be an isometry
unless $\<\W,\W\>=1.$
\er

\begin{physics}
The map $m_{\mA}\big((\pi,\Hi,\W)\big)$ tells us
that if we start with an arbitrary representation
$(\pi,\Hi)$ of the algebra of observables $\mA$
together with a normalized 
vector $\W\in\Hi$ (the representation
need not need be irreducible and the vector
need not be cyclic), if we forget about our
Hilbert space, and remember only the algebra
of observables and our state, then we might not
be able to recover our exact Hilbert space back,
but we can get close. The best we can do from the 
GNS construction is to get a new Hilbert space
that embeds into the Hilbert space we started with
via $m_{\mA}\big((\pi,\Hi,\W)\big).$ 
Furthermore, in this subspace, the vector
we started with becomes cyclic with respect to the
algebra of observables. 
In other words, we lose some information,
namely the vectors orthogonal to this subspace,
but we keep many of the essential features of our
initial state and our algebra of observables. 
\end{physics}

\blem
\label{lem:m}
$m$ from Construction \ref{con:modification}
defines a modification (cf. Definition
\ref{defn:GNSmodification})
\be
\xy0;/r.25pc/:
(-15,-7.5)*+{\pRep}="1";
(0,7.5)*+{\St}="2";
(15,-7.5)*+{\pRep}="3";
{\ar@{=>}"1";"2"^{\rest}};
{\ar@{=>}"2";"3"^{\pGNS}};
{\ar@{=>}"1";"3"_{\id_{\pRep}}};
{\ar@3{->}(0,3);(0,-6)_{m}};
\endxy
\ee
via the assignment
\be
\CAlg_{0}\ni\mA
\xmapsto{\qquad}
\quad
\xy0;/r.25pc/:
(-15,-7.5)*+{\pRep(\mA)}="1";
(0,7.5)*+{\St(\mA)}="2";
(15,-7.5)*+{\pRep(\mA)}="3";
{\ar"1";"2"^{\rest_{\mA}}};
{\ar"2";"3"^{\pGNS_{\mA}}};
{\ar"1";"3"_{\id_{\pRep(\mA)}}};
{\ar@{=>}"2";(0,-6.5)|-(0.55){m_{\mA}}};
\endxy
.
\ee
Furthermore, for each $C^*$-algebra $\mA,$
$m_{\mA}$ restricts to a well-defined and
vertically invertible natural transformation on the subcategory $\cRep(\mA).$
\elem

\bprf
In order for $m$ to be a modification, for every 
morphism $f:\mA'\to\mA$ of $C^*$-algebras,
the following equality must hold (see equation (\ref{eq:coherencemodification}))
\be
\xy0;/r.24pc/:
(25,10)*+{\pRep(\mA)}="1";
(-25,10)*+{\pRep(\mA)}="2";
(25,-10)*+{\pRep(\mA')}="3";
(-25,-10)*+{\pRep(\mA')}="4";
(0,5)*+{\St(\mA)}="5";
(0,-15)*+{\St(\mA')}="6";
{\ar@/_0.5pc/"5";"1"^(0.505){\pGNS_{\mA}}};
{\ar@/_0.5pc/"6";"3"_(0.60){\pGNS_{\mA'}}};
{\ar"1";"3"|-{\pRep(f)}};
{\ar"2";"4"|-{\pRep(f)}};
{\ar@/_0.5pc/"2";"5"^(0.45){\rest_{\mA}}};
{\ar@/_0.5pc/"4";"6"_(0.40){\rest_{\mA'}}};
{\ar"5";"6"|-{\St(f)}};
{\ar@/^1.5pc/"2";"1"^{\id_{\pRep(\mA)}}};
{\ar@{=>}"6"+(7,5);"1"+(-6,-6)|-{\pGNS_{f}}};
{\ar@{=}"4"+(7,4);"5"+(-6,-4)|-{\id=\rest_{f}}};
{\ar@{=>}"5";(0,15)|-(0.4){m_{\mA}}};
\endxy
\hspace{-2mm}
=
\hspace{-2mm}
\xy0;/r.24pc/:
(-25,-10)*+{\pRep(\mA')}="1";
(0,-15)*+{\St(\mA)}="2";
(25,-10)*+{\pRep(\mA')}="3";
(-25,10)*+{\pRep(\mA)}="4";
(25,10)*+{\pRep(\mA)}="5";
{\ar@/_0.5pc/"1";"2"_(0.40){\rest_{\mA'}}};
{\ar@/_0.5pc/"2";"3"_(0.60){\pGNS_{\mA'}}};
{\ar"4";"1"|-{\pRep(f)}};
{\ar"5";"3"|-{\pRep(f)}};
{\ar@/^1.5pc/"4";"5"^{\id_{\pRep(\mA)}}};
{\ar@/^1.5pc/"1";"3"|-{\id_{\pRep(\mA')}}};
{\ar@{=}@/^0.5pc/"1"+(7,6);"5"+(-9,-4)|-{\id_{\pRep(f)}}};
{\ar@{=>}"2";(0,-6)|-(0.4){m_{\mA'}}};
\endxy
\ee
i.e. for every object $(\pi,\Hi,\W)$
of $\pRep(\mA)$ with 
$\w:=\<\W,\pi(\;\cdot\;)\W\>,$
the diagram
\be
\label{eq:modificationGNS}
\xy0;/r.25pc/:
(-30,-7.5)*+{\big(\pi_{\w\circ f},
\Hi_{\w\circ f},[1_{\mA'}]\big)}="1";
(0,7.5)*+{\big(\pi_{\w}\circ f,
\Hi_{\w},[1_{\mA}]\big)}="2";
(30,-7.5)*+{(\pi\circ f,\Hi,\W)}="3";
{\ar"1";"2"^(0.4){\pGNS_{f}(\w)}};
{\ar"2";"3"^(0.65){\;\;f^*\left(m_{\mA}
\big((\pi,\Hi,\W)\big)\right)}};
{\ar"1";"3"_{m_{\mA'}\big((\pi\circ f,\Hi,\W)\big)}};
\endxy
\ee
of intertwiners of pointed representations of
$\mA'$ must commute. 
The image of a vector 
$[a']\in\mA'/\mathcal{N}_{\w\circ f}$ under the top
two linear maps is $\pi\big(f(a')\big)\W$ while
the image under the bottom map is
$(\pi\circ f)(a')\W.$ These are equal
elements in $\Hi.$ Because the maps
agree on a dense subspace, the diagram 
(\ref{eq:modificationGNS}) commutes. 
Finally, when $m_{\mA}$ is restricted to
the subcategory 
$\cRep(\mA),$
it was shown at the end of Construction
(\ref{con:modification})
that it is unitary on objects of $\cRep(\mA)$ 
and therefore a vertically invertible
natural transformation. 
\eprf

\begin{physics}
Commutativity of (\ref{eq:modificationGNS}), i.e. $m$ being a modification, 
encodes the fact that the way in which our GNS Hilbert spaces
sit inside the original space agrees with respect to the smaller
collection of observables in the case that one is restricting
to a subalgebra. 
\end{physics}

Everything we have done up to this point
leads to the following theorem
encompassing the GNS construction. 
To state it, we introduce the functor 2-category 
(see Definition \ref{defn:functor2cat}).

\bd
Let $\mathbf{Fun}(\CAlg^{\op},\Cat)$
be the 2-category whose objects are 
functors from $\CAlg^{\op}$ to $\Cat,$
1-morphisms are oplax-natural
transformations, and 2-morphisms
are modifications. Compositions and identities
are defined as in ordinary 2-category theory 
(see Appendix \ref{appendix} for definitions).
\ed

\bt
\label{thm:GNSmain}
\sloppy The oplax-natural
transformation
$\pGNS:\St\Rightarrow\pRep$
is left adjoint to $\rest.$ In fact,
the quadruple $(\pGNS,\rest,\id,m)$
is an adjunction in 
$\mathbf{Fun}(\CAlg^{\op},\Cat).$
\et

\bprf
The only thing left to check are the 
zig-zag identities from Lemma
\ref{lem:semipseudonaturaladjunction}.
Using the notation from that lemma,
$F:=\St,$
$G:=\pRep,$
$\s:=\pGNS,$
$\rho:=\rest,$
$\h:=\id,$
and
$\e:=m.$
By Remark \ref{rmk:zigzag},
it suffices to prove
\be
\label{eq:GNSadjunction1}
\hspace{-4mm}
\xy0;/r.25pc/:
(0,22.5)*+{\pRep(\mA)}="1";
(0,7.5)*+{\St(\mA)}="2";
(0,-7.5)*+{\pRep(\mA)}="3";
(0,-22.5)*+{\St(\mA)}="4";
{\ar"1";"2"|-{\rest_{\mA}}};
{\ar"2";"3"|-{\pGNS_{\mA}}};
{\ar"3";"4"|-{\rest_{\mA}}};
{\ar@/_4.5pc/"2";"4"_{\id_{\St(\mA)}}};
{\ar@/^4.5pc/"1";"3"^{\id_{\pRep(\mA)}}};
{\ar@2{->}(-17.5,-7.5);"3"^(0.25){\id}};
{\ar@2{->}"2";(17.5,7.5)^(0.75){m_{\mA}}};
\endxy
\
=
\quad
\xy0;/r.25pc/:
(0,10)*+{\pRep(\mA)}="1";
(0,-10)*+{\St(\mA)}="2";
{\ar@/_2.25pc/"1";"2"_{\rest_{\mA}}};
{\ar@/^2.25pc/"1";"2"^{\rest_{\mA}}};
{\ar@2{->}(-7.5,0);(7.5,0)^{\id_{\rest_{\mA}}}};
\endxy
\ee
and
\be
\label{eq:GNSadjunction2}
\hspace{-4mm}
\xy0;/r.25pc/:
(0,22.5)*+{\St(\mA)}="1";
(0,7.5)*+{\pRep(\mA)}="2";
(0,-7.5)*+{\St(\mA)}="3";
(0,-22.5)*+{\pRep(\mA)}="4";
{\ar"1";"2"|-{\pGNS_{\mA}}};
{\ar"2";"3"|-{\rest_{\mA}}};
{\ar"3";"4"|-{\pGNS_{\mA}}};
{\ar@/^4.5pc/"2";"4"^{\id_{\pRep(\mA)}}};
{\ar@/_4.5pc/"1";"3"_{\id_{\St(\mA)}}};
{\ar@2{->}(-17.5,7.5);"2"^(0.25){\id}};
{\ar@2{->}"3";(17.5,-7.5)^(0.75){m_{\mA}}};
\endxy
\hspace{-2mm}
=
\quad
\xy0;/r.25pc/:
(0,10)*+{\St(\mA)}="1";
(0,-10)*+{\pRep(\mA)}="2";
{\ar@/_2.25pc/"1";"2"_{\pGNS_{\mA}}};
{\ar@/^2.25pc/"1";"2"^{\pGNS_{\mA}}};
{\ar@2{->}(-7.5,0);(7.5,0)^{\id_{\pGNS_{\mA}}}};
\endxy
\ee
for each object $\mA$ of $\CAlg^{\op}.$ Fortunately,
these identities are essentially tautologous.
For (\ref{eq:GNSadjunction1}), since
$\St(\mA)$ has no non-trivial morphisms,
the equality holds. For (\ref{eq:GNSadjunction2}),
it suffices to check what happens to a state $\w.$
Under the composition in (\ref{eq:GNSadjunction2}),
$\w$ gets sent to
\be
\w\mapsto
\big(\pi_{\w},\Hi_{\w},[1_{\mA}]\big)\mapsto
\big\<[1_{\mA}],\pi_{\w}(\ \cdot \ )[1_{\mA}]\big\>=\w
\mapsto\big(\pi_{\w},\Hi_{\w},[1_{\mA}]\big),
\ee
which is exactly the same representation as
in the second step. Finally, 
$m_{\mA}\big((\pi_{\w},\Hi_{\w},[1_{\mA}])\big)$
is the identity intertwiner because it sends $[a]\in\mA/\mathcal{N}_{\w}$
to $\pi(a)[1_{\mA}]$ by (\ref{eq:Segalmap}) but $\pi(a)[1_{\mA}]=[a]$ 
by (\ref{eq:GNSrepn}). 
\eprf

In particular, by Remark \ref{rmk:zigzag},
this theorem implies the following.

\bc
For every $C^*$-algebra $\mA,$ the quadruple $\big(\pGNS_{\mA},\rest_{\mA},
\id,m_{\mA}\big)$ is an adjunction (in the usual sense). 
\ec

This means that for every state $\w\in\St(\mA)_{0}$ and pointed representation 
$(\pi,\Hi,\W)\in\pRep(\mA)_{0},$ there is a natural bijection of morphisms%
\footnote{This is how one remembers that $\pGNS$ is \emph{left} adjoint
to $\rest.$}
\be
\hspace{-2mm}
\pRep(\mA)\Big(\pGNS_{\mA}(\w),(\pi,\Hi,\W)\Big)
\cong
\St(\mA)\Big(\w,\rest_{\mA}(\pi,\Hi,\W)\Big),
\ee
which illustrates 
in what sense the GNS
construction $\pGNS_{\mA}(\w)$ is optimal:
for every other choice of representation
$(\pi,\Hi,\W)$ on which to realize the state $\w$ as the restriction of a
vector state, there is always a (unique)
isometric intertwiner from the GNS Hilbert
space to $\Hi.$ In particular,
the GNS Hilbert space is the 
\emph{smallest} space on which one can
represent a state as the restriction of a vector state. 
If $\w$ and $\rest_{\mA}\big((\pi,\Hi,\W)\big)$ do not
agree on $\mA,$ this result also says that there is \emph{no} intertwiner
between the pointed representations 
$\pGNS_{\mA}(\w)$ and $(\pi,\Hi,\W)$
(since $\St(\mA)$ is a discrete category).
This can change if we have an
isomorphism of our algebra.

A special case of the adjunction
$\big(\pGNS_{\mA},\rest_{\mA},
\id,m_{\mA}\big)$ 
occurs when restricted
to the category of cyclic representations
of $\mA.$ In this case, 
it is an adjoint equivalence (meaning, an
equivalence of categories). 
In other words, in the cyclic case,
the categories $\cRep(\mA)$
and $\St(\mA)$ are equivalent
and the restriction functor exhibits this equivalence
with a canonical inverse given by the
GNS construction. In particular,
this reproduces the well-known result \cite{Se47} that
there is a one-to-one correspondence between
isomorphism classes of cyclic representations
of $\mA$ and states on $\mA.$

Our results can be summarized
by saying that we can now provide
a \emph{definition}
instead of a \emph{construction}
that produces, in a functorial manner,
cyclic representations from states
on $C^*$-algebras.

\bd
\label{defn:GNS}
The \emph{\uline{GNS construction}}
is the left adjoint of $\rest.$ 
\ed

This means that the GNS construction is characterized by 
the following data: 
\begin{enumerate}[i)]
\item
a function
$
\mG_{\mA}:\mS(\mA)\to\pRep(\mA)_{0}
$
for each $C^*$-algebra $\mA$
sending states on $\mA$ to pointed representations of $\mA,$

\item
a function 
$
\mG_{f}:\mS(\mA)\to\pRep(\mA')_{1}
$
for each morphism $f:\mA'\to\mA$ of $C^*$-algebras
sending states on $\mA$ to isometric intertwiners of pointed 
representations of $\mA',$ and

\item
a function
$
m_{\mA}:\pRep(\mA)_{0}\to\pRep(\mA)_{1}
$
for each $C^*$-algebra $\mA$ 
\end{enumerate}
subject to the following conditions: 
\begin{enumerate}[(a)]
\item
$\mG_{f}(\w)$ is an isometric intertwiner 
of representations of $\mA'$ from
$\mG_{\mA'}(\w\circ f)$ to%
\footnote{Here, $f^*$ is the pullback of a representation 
along the map $f.$}
 $f^*\mG_{\mA}(\w)$
for all states $\w\in\mS(\mA),$

\item
$m_{\mA}\big((\pi,\Hi,\W)\big)$ is a isometric intertwiner 
from $\mG_{\mA}\Big(\rest_{\mA}\big((\pi,\Hi,\W)\big)\Big)$ 
to $(\pi,\Hi,\W)$ for all pointed representations 
$(\pi,\Hi,\W)\in\pRep(\mA)_{0},$

\item
$\mG_{\id_{\mA}}=i_{\mA}\circ\mG_{\mA},$
where $i_{\mA}:\pRep(\mA)_{0}\to\pRep(\mA)_{1}$ is the
map that assigns the identity intertwiner to each pointed
representation, 

\item
$\mG_{f\circ f'}(\w)=f'^{*}\big(\mG_{f}(\w)\big)\circ\mG_{f'}(\w\circ f)$
for all composable pairs of $C^*$-algebra morphisms 
$\mA''\xrightarrow{f'}\mA'\xrightarrow{f}\mA$ and states
$\w\in\mS(\mA),$ 

\item
$f^*m_{\mA}\big((\pi,\Hi,\W)\big)\circ
\mG_{f}\Big(\rest_{\mA}\big((\pi,\Hi,\W)\big)\Big)
=m_{\mA'}\big((\pi\circ f,\Hi,\W)\big)$ for all
pointed representations $(\pi,\Hi,\W)\in\pRep(\mA)_{0},$

\item
$\rest_{\mA}\circ\mG_{\mA}=\id_{\mS(\mA)},$ and

\item
$m_{\mA}\big(\mG_{\mA}(\w)\big)=\id_{\mG_{\mA}(\w)}$
for all states $\w\in\mS(\mA).$ 

\end{enumerate}

The reader should compare this short definition to Constructions 
\ref{con:GNSobjects}, \ref{con:GNSmorphisms}, and
\ref{con:GNScyclicpseudo}, which are
the usual definitions of the GNS construction. 
Many of these properties are well-known, though less emphasis 
is placed on allowing non-unitary intertwiners, 
obfuscating the underlying categorical perspective outlined here. 
We have established such a categorical framework
for the GNS constructing viewing it as a morphism/process
in an appropriate category, namely $\mathbf{Fun}(\CAlg^{\op},\Cat),$
and we have illustrated that the universal properties 
of the GNS construction are encoded in this morphism being
left adjoint to the simple morphism/process that views normalized
vectors as states.

\section{Examples}
\label{sec:GNSex}

The authors
of \cite{BGdQRL} include several examples, and we
will go through the simplest ones to illustrate the 
meaning of our constructions and theorems. 

\begin{example}
Let $\mA:=\mB(\C^2),$ $2\times2$
matrices with complex coefficients. 
This is the algebra of observables for a 
spin-$\frac{1}{2}$ system, i.e. a qubit.
Label an orthonormal basis by 
$\{\up,\dn\}$---this basis refers to the 
spin of a particle
along a particular axis.
Let $\mA$ act on $\C^2$ by the identity
representation, meaning that the representation
$\pi:\mA\to\mB(\C^2)$ is just the identity map. 
Let $\w_{\up}:\mB(\C^2)\to\C$ be the state
corresponding to a pure state with spin up, i.e.%
\footnote{We are using Dirac bra-ket notation
for the examples.}
$\w_{\up}(a):=\<\up\!|a|\!\up\>$ for all $a\in\mB(\Hi).$
Applying the restriction functor $\rest_{\mA}$
to the pointed representation
$\big(\pi,\C^2,|\!\!\up\>\big)$ gives $\w_{\up}.$
Next, apply the GNS construction $\pGNS_{\mA}$
to the state $\w_{\up}.$ As a vector space, 
$\mB(\C^2)$ is four-dimensional
with a basis given by
\be
\label{ex:eik}
e_{\up\up}=\begin{pmatrix}1&0\\0&0\end{pmatrix},
\quad
e_{\up\dn}=\begin{pmatrix}0&1\\0&0\end{pmatrix},
\quad
e_{\dn\up}=\begin{pmatrix}0&0\\1&0\end{pmatrix},
\quad\&\quad
e_{\dn\dn}=\begin{pmatrix}0&0\\0&1\end{pmatrix}.
\ee
The expectation values for these operators are
given by
\be
\w_{\up}(e_{\up\up})=1,
\quad
\w_{\up}(e_{\up\dn})=0,
\quad
\w_{\up}(e_{\dn\up})=0,
\quad\&\quad
\w_{\up}(e_{\dn\dn})=0.
\ee
Notice that%
\footnote{To avoid confusion with
the physics literature,
for the purposes of this section,
we will use ${}^{\dagger}$ to denote
the adjoint instead of ${}^*.$}
$e_{\up\dn}^{\dag}e_{\up\dn}=e_{\dn\dn}$
and $e_{\dn\dn}^{\dag}e_{\dn\dn}=e_{\dn\dn}$ so that
$\w_{\up}(e_{\up\dn}^{\dag}e_{\up\dn})=0$
and $\w_{\up}(e_{\dn\dn}^{\dag}e_{\dn\dn})=0.$
In fact, 
\be
\mathcal{N}_{\w_{\up}}
=\mathrm{span}\{e_{\up\dn},e_{\dn\dn}\}.
\ee
Then,%
\footnote{No completion is necessary here since the vector spaces
are finite-dimensional.}
$\Hi_{\w_{\up}}:=\mB(\C^2)/\mathcal{N}_{\w_{\up}}$
consists of equivalence classes of matrices
\be
a=
\begin{pmatrix}
a_{\up\up}&a_{\up\dn}\\
a_{\dn\up}&a_{\dn\dn}
\end{pmatrix}
,
\ee
where $a_{ij}\in\C$ with $i,j\in\{\up,\dn\},$ 
and $a\sim b$ if and only if
\be
b-a=\begin{pmatrix}0&b_{\up\dn}-a_{\up\dn}\\
0&b_{\dn\dn}-a_{\dn\dn}\end{pmatrix}
.
\ee
Note that the inner product on $\Hi_{\w_{\up}}$ in this case is given by 
\be
\Hi_{\w_{\up}}\times\Hi_{\w_{\up}}\ni([a],[b])\mapsto\w_{\up}(a^{\dag}b)
=\overline{a_{\up\up}}b_{\up\up}+\overline{a_{\dn\up}}b_{\dn\up},
\ee
where the overline denotes complex conjugation.
The associated cyclic representation from the 
GNS construction applied to the state $\w_{\up}$ is
$(\pi_{\w_{\up}},\Hi_{\w_{\up}},[\mathds{1}]),$
where $\mathds{1}$ is the $2\times 2$ identity
matrix and $\pi_{\w_{\up}}(a)\big([b]\big)=[ab]$
is obtained from ordinary matrix multiplication.
The intertwiner $m_{\mA}$ from (\ref{eq:Segalmap})
applied to our representation $\big(\pi,\C^2,|\!\up\>\big)$
is the map $[a]\mapsto a|\!\!\up\>.$ Since our representation
was cyclic to begin with, this intertwiner is unitary. This
map compares our original Hilbert space representation
to the one obtained from the GNS construction in a
canonical way. 
\end{example}

\begin{example}
Let $\mA:=\mB(\C^2)\otimes\mB(\C^2)$
and let $\mA$ act on $\C^2\otimes\C^2$ via 
\be
\label{eq:EPRpi}
\begin{split}
\mA&\xrightarrow{\pi}\mB(\C^2\otimes\C^2)\\
a\otimes b&\mapsto\Big(\C^2\otimes\C^2\ni|\psi\>\otimes|\phi\>\xmapsto{\pi(a\otimes b)} a|\psi\>\otimes b|\phi\>\Big)
\end{split}
\ee
on elementary tensors and appropriately extended. 
Let $\w$ be the state 
\be
\mA\ni a\otimes b\mapsto \w(a\otimes b):=\<\Psi|\pi(a\otimes b)|\Psi\>
\ee
on $\mA$ obtained by pulling back 
the vector state associated to the unit vector
\be
\label{eq:EPRstate}
|\Psi\>:=
\frac{1}{\sqrt{2}}
\Big(|\!\up\dn\>-|\!\dn\up\>\Big),
\ee
where $|\hspace{-1mm}\up\dn\>$ is short for $|\hspace{-1mm}\up\>\otimes|\hspace{-1mm}\dn\>.$
Let the inner product on $\C^2\otimes\C^2$
be the usual Euclidean inner product.
Using the same notation as in the previous example,
\be
\begin{split}
\w(a\otimes b)&=\frac{1}{2}
\Big(\<\up\dn\!|-\<\dn\up\!|\Big)
\pi(a\otimes b)\Big(|\!\up\dn\>-|\!\dn\up\>\Big)\\
&=\frac{1}{2}\Big(\<\up\!|\otimes\<\dn\!|-\<\dn\!|\otimes\<\up\!|\Big)\Big(a|\!\up\>\otimes b|\!\dn\>-a|\!\dn\>\otimes b|\!\up\>\Big)\\
&=\frac{1}{2}
\Big(a_{\up\up}b_{\dn\dn}-a_{\up\dn}b_{\dn\up}
-a_{\dn\up}b_{\up\dn}+a_{\dn\dn}b_{\up\up}\Big).
\end{split}
\ee
It is not necessary for us to calculate
$\Hi_{\w}$ explicitly but we use the isomorphism $\pi$ in (\ref{eq:EPRpi})
to identify its elements as 
equivalence classes $[a\otimes b]$ 
of elements in $\mB(\C^2)\otimes\mB(\C^2).$ 
Because $|\Psi\>$ is cyclic, 
Lemma \ref{lem:m} shows that the map
\be
\label{eq:mAsu2intertwiner}
\begin{split}
\big(\pi_{\w},\Hi_{\w},[\mathds{1}]\big)
&\xrightarrow{m_{\mB(\C^2\otimes\C^2)}}
\big(\pi,\C^2\otimes\C^2,|\Psi\>\big)\\
[a\otimes b]&\xmapsto{\qquad\quad\;\;\;}
\pi(a\otimes b)|\Psi\>
\end{split}
\ee
extended linearly is a unitary intertwiner. 
Now, let $i_1:\mB(\C^2)\to\mB(\C^2)\otimes\mB(\C^2)$
be the $C^*$-algebra map defined by
\be
i_1(a):=a\otimes\mathds{1}.
\ee
Physically, such a map corresponds to an
observer $\mathcal{O}_1$
only being able to make measurements
on the observables $\mB(\C^2)$ 
corresponding to a single particle. It is convenient
to denote the first $\C^2$ by $\Hi_{1}$
and the second by $\Hi_{2}.$
This situation occurs, for instance, 
in an EPR-like experiment,
where a particle decomposes into two particles whose spins
are correlated in a way described by the vector (\ref{eq:EPRstate}).
The two particles fly off in opposite directions and
observers far away are waiting to measure the spin. 
\[
\xy0;/r.25pc/:
(-50,0)*+{\mathcal{O}_{1}}="1";
(0,0)*+{}="w";
(0,5)*+{\frac{1}{\sqrt{2}}
\Big(|\!\up\dn\>-|\!\dn\up\>\Big)};
(50,0)*+{\mathcal{O}_{2}}="2";
{\ar@{~>}"w";"1"};
{\ar@{~>}"w";"2"};
\endxy
\]
Observer $\mathcal{O}_{1}$ cannot measure
the observables $\mB(\Hi_{2})$ and vice versa.
Therefore, the state that $\mathcal{O}_{1}$
sees is given by the restriction 
\be
\w_{1}:=\w\circ i_{1}:\mB(\Hi_1)\to
\mB(\Hi_1)\otimes\mB(\Hi_2)\to\C.
\ee
This state corresponds to the density matrix
\be
\rho_{1}=\begin{pmatrix}\frac{1}{2}&0\\
0&\frac{1}{2}\end{pmatrix}
\ee
on $\Hi_{1}$ using our ordered basis 
$({|\hspace{-1mm}\up\>},{|\hspace{-1mm}\dn\>}).$
What is the GNS construction
applied to the state $\w_{1}$ and how is it related to 
the original pointed representation 
$\big(\pi,\Hi_{1}\otimes\Hi_{2},|\Psi\>\big)$?
Let $a\in\mB(\Hi_1).$ Then
\be
\big(a^{\dag}a\big)_{ik}
=\sum_{j\in\{\up,\dn\}}(a^{\dag})_{ij}a_{jk}
=\sum_{j\in\{\up,\dn\}}\overline{a_{ji}}a_{jk}
\ee
implies
\be
\begin{split}
\w_{1}(a^{\dag}a)&=\tr(\rho_{1}a^{\dag}a)
=\frac{1}{2}\Big(\<\up\!|a^{\dag}a|\!\up\>
+\<\dn\!|a^{\dag}a|\!\dn\>\Big)\\
&=\frac{1}{2}\sum_{j\in\{\up,\dn\}}\Big(
|a_{j\up}|^2+|a_{j\dn}|^2\Big)
=\frac{1}{2}\sum_{j,k\in\{\up,\dn\}}|a_{jk}|^2
\end{split}
\ee
so that $\w_{1}(a^{\dag}a)=0$
if and only if $a=0.$ Therefore, $\mathcal{N}_{\w_1}=\{0\}$
and hence $\Hi_{\w_1}=\mB(\Hi_{1})$ as a vector space.
The inner product on $\Hi_{\w_{1}}$ is given by 
\be
\Hi_{\w_1}\times\Hi_{\w_1}\ni(a,b)\mapsto\w_{1}(a^{\dag}b)=\frac{1}{2}\tr\left(a^{\dag}b\right),
\ee
which is half the Hilbert-Schmidt (Frobenius) inner product. 
Furthermore, 
the associated GNS representation $\pi_{\w_1}$
acts as
\be
\pi_{\w_1}(a)b
=ab
=\sum_{i,j,k\in\{\up,\dn\}}a_{ij}b_{jk}e_{ik},
\ee
where the $e_{ik}$ are as in (\ref{ex:eik}). 
The induced map $L_{i_1}:\Hi_{\w_{1}}\to\Hi_{\w}$
corresponding to
(\ref{eq:eq:GNSintertwinerdiagram}) is
given by 
\be
\begin{split}
\mB(\Hi_1)\equiv
\Hi_{\w_{1}}&\xrightarrow{L_{i_1}=\pGNS_{i_1}(\w)}
\Hi_{\w}\\
a&\xmapsto{\qquad\qquad\quad}[a\otimes\mathds{1}]
\end{split}
\ee
Using this with the intertwiner 
$m_{\mB(\Hi_1\otimes\Hi_2)}$
from (\ref{eq:mAsu2intertwiner}),
gives a canonical intertwiner
of $\mB(\Hi_{1})$-representations
to our original Hilbert space
\be
\label{eq:exampleSU2intertwiner}
\xy0;/r.25pc/:
(-35,3)*+{\Hi_{\w_{1}}}="1";
(0,3)*+{\Hi_{\w}}="2";
(35,3)*+{\Hi_{1}\otimes\Hi_{2}}="3";
(-35,-3)*+{a}="a";
(0,-3)*+{[a\otimes\mathds{1}]}="aotimes1";
(35,-3)*+{\pi(a\otimes\mathds{1})|\Psi\>}="a1";
{\ar"1";"2"^{\GNS^{\bullet}_{i_1}(\w)}};
{\ar"2";"3"^(0.4){i_{1}^{*}(m_{\mB(\Hi_1\otimes\Hi_2)})}};
{\ar@{|->}"a";"aotimes1"};
{\ar@{|->}"aotimes1";"a1"};
\endxy
.
\ee
This canonical map is the top arrow in
the diagram (\ref{eq:modificationGNS}).
This exhibits our Hilbert space
$\Hi_{\w_1},$ which was the Hilbert space
from the GNS construction associated
to the EPR density matrix $\rho_{1}$ for observer $\mathcal{O}_{1},$
as a vector subspace of our original Hilbert
space $\Hi_{1}\otimes\Hi_{2}$ 
for the entangled EPR vector $|\Psi\>.$
Note that the map (\ref{eq:exampleSU2intertwiner})
is surjective because
\be
\pi(a\otimes\mathds{1})|\Psi\>
=\frac{1}{\sqrt{2}}\Big(a_{\up\up}|\!\up\dn\>+a_{\dn\up}|\!\dn\dn\>
-a_{\up\dn}|\!\up\up\>-a_{\dn\dn}|\!\dn\up\>\Big).
\ee
It is also an isometry by Lemma \ref{lem:isometry} because
\be
\lVert\pi(a\otimes\mathds{1})|\Psi\>\rVert^2
=\<\Psi|\pi(a^{\dag}\otimes\mathds{1})\pi(a\otimes\mathds{1})|\Psi\>
=\frac{1}{2}\sum_{j,k\in\{\up,\dn\}}\!|a_{jk}|^2=\frac{1}{2}\tr(a^{\dag}a)
=\lVert a\rVert_{\w_{1}}^{2}
\ee
for all $a\in\Hi_{\w_{1}}.$ 
Hence, (\ref{eq:exampleSU2intertwiner}) is a unitary intertwiner.
\end{example}

\appendix
\section{2-categorical preliminaries}
\label{appendix}

In the GNS construction, we use 
oplax-natural transformations,
which are different from the
pseudo-natural transformations
that appear in the early literature on 2-categories \cite{Be}. 
Fortunately, the difference is minor.
For completeness, we include this definition
along with the notion of modifications \cite{Bo94}. 

\bd
\label{defn:semipseudonaturaltransformation}
Let $\mathcal{C}$ and $\mathcal{D}$ be two
(strict)%
\footnote{A definition exists for weak 2-categories
and weak 2-functors but such a definition is
not needed here.}
2-categories and let $F, G : \mathcal{C} \to \mathcal{D}$ 
be two 2-functors. An
\emph{\uline{oplax-natural transformation}}
$\rho$ from $F$ to $G$, written as 
$\rho:F\Rightarrow G,$ consists of
\begin{enumerate}[i)]
\item
a function $\rho : C_{0} \to D_{1}$ assigning a 
1-morphism in $\mD$ to an object $x$ in $\mC$ in the following manner 
\be
\xy0;/r.15pc/:
(-30,0)*+{x}="1";
(30,10)*+{F(x)}="2";
(30,-10)*+{G(x)}="3";
{\ar^{\rho(x)} "2";"3"};
{\ar@{|->}^{\rho} "1"+(20,0);(10,0)};
\endxy
\ee

\item
and a function $\rho : C_{1} \to D_{2}$ assigning 
a 2-morphism%
\footnote{For a pseudo-natural transformation,
one requires this 2-morphism
to be vertically invertible, motivated by the
fact that equations should replace 
isomorphisms upon categorification
\cite{BD}.
In cases where invertibility is not imposed, one
has two possibilities depending on the 
direction of this 2-morphism. A lax-natural transformation
(see Definition 7.5.2 of \cite{Bo94}) uses the opposite
direction, which is why we use the prefix oplax. 
}
in $\mD$ to every 
1-morphism $y \xleftarrow{\a} x$ in $\mC$ in the following manner 
\be
\xy0;/r.15pc/:
(-30,0)*+{x}="1x";
(-60,0)*+{y}="1y";
{\ar_{\a} "1x";"1y"};
(30,10)*+{F(y)}="2y";
(30,-10)*+{G(y)}="3y";
(60,10)*+{F(x)}="2x";
(60,-10)*+{G(x)}="3x";
{\ar^{\rho(x)} "2x";"3x"};
{\ar_{\rho(y)} "2y";"3y"};
{\ar_{F(\a)} "2x";"2y"};
{\ar^{G(\a)} "3x";"3y"};
{\ar@{=>}^{\rho(\a)} "2y";"3x"};
{\ar@{|->}^{\rho} "1x"+(20,0);(10,0)};
\endxy
.
\ee
\end{enumerate}

These data must satisfy the following conditions:

\begin{enumerate}[(a)]
\item
For every object $x$ in $\mC,$ 
\be
\rho(\id_{x})=\id_{\rho(x)}.
\ee

\item
For every pair $(z\xleftarrow{\a}y,y\xleftarrow{\b}x)$ 
of composable 1-morphisms in $\mathcal{C},$ the 
diagram%
\footnote{Horizontal composition of 2-morphisms is 
written using $\circ.$}
\be
\xy 0;/r.15pc/:
(-42,20)*+{\rho(z)\circ F(\a)\circ F(\b)}="1";
(42,20)*+{G(\a)\circ\rho(y)\circ F(\b)}="2R";
(42,-20)*+{G(\a)\circ G(\b)\circ\rho(x)}="3R";
(-42,0)*+{\rho(z)\circ F(\a\b)}="2L";
(-42,-20)*+{G(\a\b)\circ\rho(x)}="3L";
{\ar@2{->}"1";"2R"^{\rho(\a)\circ\id_{F(\b)}}};
{\ar@2{->}"2R";"3R"^{\id_{G(\a)}\circ\rho(\b)}};
{\ar@{=}"1";"2L"_{\id}};
{\ar@2{->}"2L";"3L"_{\rho(\a\b)}};
{\ar@{=}"3L";"3R"_{\id}};
\endxy
\ee
commutes, i.e. 
\be
\xy0;/r.20pc/:
(-30,10)*+{F(z)}="2z";
(-30,-10)*+{G(z)}="3z";
(0,10)*+{F(y)}="2y";
(0,-10)*+{G(y)}="3y";
(30,10)*+{F(x)}="2x";
(30,-10)*+{G(x)}="3x";
{\ar|-{\rho(x)} "2x";"3x"};
{\ar|-{\rho(y)} "2y";"3y"};
{\ar|-{\rho(z)} "2z";"3z"};
{\ar_{F(\b)} "2x";"2y"};
{\ar^{G(\b)} "3x";"3y"};
{\ar_{F(\a)} "2y";"2z"};
{\ar^{G(\a)} "3y";"3z"};
{\ar@{=>}^{\rho(\b)} "2y";"3x"};
{\ar@{=>}^{\rho(\a)} "2z";"3y"};
\endxy
\
=
\
\xy0;/r.20pc/:
(-15,10)*+{F(z)}="2z";
(-15,-10)*+{G(z)}="3z";
(15,10)*+{F(x)}="2x";
(15,-10)*+{G(x)}="3x";
{\ar|-{\rho(x)} "2x";"3x"};
{\ar|-{\rho(z)} "2z";"3z"};
{\ar_{F(\a\b)} "2x";"2z"};
{\ar^{G(\a\b)} "3x";"3z"};
{\ar@{=>}^{\rho(\a\b)} "2z";"3x"};
\endxy
.
\ee
\item
For every 2-morphism 
\be
\xymatrix{
y &&
\ar@/_2pc/[ll]_{\a}="8"
\ar@/^2pc/[ll]^{\g}="9"
\ar@{}"8";"9"|(.2){\,}="10"
\ar@{}"8";"9"|(.8){\,}="11"
\ar@{=>}"10";"11"^{\S}
x 
}
,
\ee
the diagram 
\be
\xy0;/r.20pc/:
(20,10)*+{\rho(y)\circ F(\a)}="1";
(-20,10)*+{G(\a)\circ\rho(x)}="2";
(-20,-10)*+{G(\g)\circ\rho(x)}="3";
(20,-10)*+{\rho(y)\circ F(\g)}="4";
{\ar@2{->}"1";"2"_{\rho(\a)}};
{\ar@2{->}"2";"3"_{G(\S)\circ\id_{\rho(x)}}};
{\ar@2{->}"4";"3"^{\rho(\g)}};
{\ar@2{->}"1";"4"^{\id_{\rho(y)}\circ F(\S)}};
\endxy
\ee
commutes, i.e. 
\be
\xy0;/r.25pc/:
(-12.5,7.5)*+{F(y)}="Fy";
(-12.5,-7.5)*+{G(y)}="Gy";
(12.5,7.5)*+{F(x)}="Fx";
(12.5,-7.5)*+{G(x)}="Gx";
{\ar@/_1.25pc/"Fx";"Fy"_{F(\a)}};
{\ar@/^1.25pc/"Fx";"Fy"|-{F(\g)}};
{\ar@/^1.25pc/"Gx";"Gy"^{G(\g)}};
{\ar"Fy";"Gy"_{\rho(y)}};
{\ar"Fx";"Gx"^{\rho(x)}};
{\ar@{=>}@/_0.75pc/"Fy"+(3,-5);"Gx"_(0.4){\rho(\g)}};
{\ar@{=>}(0,12);(0,4)|-{F(\S)}};
\endxy
\qquad
=
\qquad
\xy0;/r.25pc/:
(-12.5,7.5)*+{F(y)}="Fy";
(-12.5,-7.5)*+{G(y)}="Gy";
(12.5,7.5)*+{F(x)}="Fx";
(12.5,-7.5)*+{G(x)}="Gx";
{\ar@/_1.25pc/"Fx";"Fy"_{F(\a)}};
{\ar@/_1.25pc/"Gx";"Gy"|-{G(\a)}};
{\ar@/^1.25pc/"Gx";"Gy"^{G(\g)}};
{\ar"Fy";"Gy"_{\rho(y)}};
{\ar"Fx";"Gx"^{\rho(x)}};
{\ar@{=>}@/^0.75pc/"Fy";"Gx"^{\rho(\a)}};
{\ar@{=>}(0,-4);(0,-12)|-{G(\S)}};
\endxy
.
\ee
\end{enumerate}
\ed

The definition of a modification does not
change if one uses oplax-natural
transformations instead of
pseudo-natural transformations. 

\bd
\label{defn:GNSmodification}
Let $\mathcal{C}$ and $\mathcal{D}$ be two 
2-categories, $F,G:\mathcal{C}\to\mathcal{D}$ 
be two 2-functors, and $\rho,\s:F\Rightarrow G$ 
be two oplax-natural transformations. A 
\emph{\uline{modification}} $m$ 
from $\s$ to $\rho$, written as 
$m:\s\Rrightarrow\rho$ and drawn as 
\be
\xy0;/r.15pc/:
(0,0)*+{\mathcal{C}}="2";
(-40,0)*+{\mathcal{D}}="3";
{\ar@/_2.25pc/_{F} "2";"3"};
{\ar@/^2.25pc/^{G} "2";"3"};
{\ar@2{->}(-30,7.5);(-30,-7.5)_{\r}};
{\ar@{=>}(-10,7.5);(-10,-7.5)^{\s}};
{\ar@3{->}(-13,0);(-27,0)_{m}};
\endxy
,
\ee
consists of a function $m:C_{0}\to D_{2}$ 
assigning a 2-morphism in $\mD$ to an object $x$ in $\mC$ in the 
following manner 
\be
\xy0;/r.15pc/:
(-30,0)*+{x}="1";
(50,15)*+{F(x)}="2";
(50,-15)*+{G(x)}="3";
{\ar@/_1.75pc/_{\rho(x)}"2";"3"};
{\ar@/^1.75pc/^{\s(x)}"2";"3"};
{\ar@2{->}(60,0);(40,0)_{m(x)}};
{\ar@{|->}^{m}"1"+(20,0);(10,0)};
\endxy
.
\ee
This assignment must satisfy the condition that 
for every 1-morphism $y\xleftarrow{\a}x,$ the diagram 
\be
\xy0;/r.20pc/:
(20,10)*+{\s(y)\circ F(\a)}="1";
(-20,10)*+{G(\a)\circ\s(x)}="2";
(20,-10)*+{\rho(y)\circ F(\a)}="3";
(-20,-10)*+{G(\a)\circ\rho(x)}="4";
{\ar@2{->}"1";"2"_{\s(\a)}};
{\ar@2{->}"2";"4"_{\id_{G(\a)}\circ m(x)}};
{\ar@2{->}"3";"4"^{\rho(\a)}};
{\ar@2{->}"1";"3"^{m(y)\circ\id_{F(\a)}}};
\endxy
\ee
commutes, i.e.
\be
\label{eq:coherencemodification}
\xy0;/r.25pc/:
(-12.5,10)*+{F(y)}="Fy";
(-12.5,-10)*+{G(y)}="Gy";
(12.5,10)*+{F(x)}="Fx";
(12.5,-10)*+{G(x)}="Gx";
{\ar"Fx";"Fy"_{F(\a)}};
{\ar"Gx";"Gy"^{G(\a)}};
{\ar@/_1.75pc/_{\s(y)}"Fy";"Gy"};
{\ar@/^1.75pc/|-{\rho(y)}"Fy";"Gy"};
{\ar@/^1.75pc/|-{\rho(x)}"Fx";"Gx"};
{\ar@2{->}(-18.5,0);(-8.5,0)^{m(y)}};
{\ar@2{->}@/^0.75pc/"Fy";"Gx"^(0.6){\rho(\a)}};
\endxy
\
=
\
\xy0;/r.25pc/:
(-12.5,10)*+{F(y)}="Fy";
(-12.5,-10)*+{G(y)}="Gy";
(12.5,10)*+{F(x)}="Fx";
(12.5,-10)*+{G(x)}="Gx";
{\ar"Fx";"Fy"_{F(\a)}};
{\ar"Gx";"Gy"^{G(\a)}};
{\ar@/_1.75pc/|-{\s(y)}"Fy";"Gy"};
{\ar@/_1.75pc/|-{\s(x)}"Fx";"Gx"};
{\ar@/^1.75pc/^{\rho(x)}"Fx";"Gx"};
{\ar@2{->}(8.5,0);(18.5,0)^{m(x)}};
{\ar@2{->}@/_0.75pc/"Fy";"Gx"_(0.35){\s(\a)}};
\endxy
.
\ee
\ed

Compositions of oplax-natural transformations
and modifications are not changed as a result
of these alterations to the usual definitions.
In particular, the vertical composition of oplax-natural
transformations is denoted using vertical concatenation as in
\be
\xy0;/r.15pc/:
(25,0)*+{\mathcal{C}}="2";
(-25,0)*+{\mathcal{D}}="3";
{\ar@/_2.25pc/_{F} "2";"3"};
{\ar@/^2.25pc/^{H} "2";"3"};
{\ar@2{->}(0,13.5);(0,-13.5)_{\begin{smallmatrix}\r\\\s\end{smallmatrix}}};
\endxy
\quad:=\quad
\xy0;/r.15pc/:
(25,0)*+{\mathcal{C}}="2";
(-25,0)*+{\mathcal{D}}="3";
{\ar@/_2.25pc/_{F} "2";"3"};
{\ar|-(0.5){G} "2";"3"};
{\ar@/^2.25pc/^{H} "2";"3"};
{\ar@2{->}(0,13.5);(0,2.5)_{\r}};
{\ar@{=>}(0,-2.5);(0,-13.5)_{\s}};
\endxy
\ee
and is defined by the assignments
\be
\xy0;/r.15pc/:
(-30,0)*+{x}="1";
(30,20)*+{F(x)}="2";
(30,0)*+{G(x)}="3";
(30,-20)*+{H(x)}="4";
{\ar^{\rho(x)} "2";"3"};
{\ar^{\s(x)} "3";"4"};
{\ar@{|->}^{\begin{smallmatrix}\r\\\s\end{smallmatrix}} "1"+(20,0);(10,0)};
\endxy
\ee
for each object $x$ in $\mC$ and 
\be
\xy0;/r.15pc/:
(-30,0)*+{x}="1x";
(-60,0)*+{y}="1y";
{\ar_{\a} "1x";"1y"};
(30,20)*+{F(y)}="2y";
(30,0)*+{G(y)}="3y";
(30,-20)*+{H(y)}="4y";
(70,20)*+{F(x)}="2x";
(70,0)*+{G(x)}="3x";
(70,-20)*+{H(x)}="4x";
{\ar@{|->}^{\rho} "1x"+(20,0);(10,0)};
{\ar_{F(\a)} "2x";"2y"};
{\ar^{\rho(x)} "2x";"3x"};
{\ar_{\rho(y)} "2y";"3y"};
{\ar|-{G(\a)} "3x";"3y"};
{\ar@{=>}^{\rho(\a)} "2y";"3x"};
{\ar^{\s(x)} "3x";"4x"};
{\ar_{\s(y)} "3y";"4y"};
{\ar^{H(\a)} "4x";"4y"};
{\ar@{=>}^{\s(\a)} "3y";"4x"};
\endxy
\ee
for each morphism $y \xleftarrow{\a} x$ in $\mC.$ 
Similarly, the vertical composition of modifications is
denoted using vertical concatenation as in 
\be
\xy0;/r.15pc/:
(27,0)*+{\mathcal{C}}="2";
(-27,0)*+{\mathcal{D}}="3";
{\ar@/_2.25pc/_{F} "2";"3"};
{\ar@/^2.25pc/^{H} "2";"3"};
{\ar@2{->}(-15,7.5);(-15,-7.5)_{\begin{smallmatrix}\r\\\l\end{smallmatrix}}};
{\ar@{=>}(15,7.5);(15,-7.5)^{\begin{smallmatrix}\s\\\t\end{smallmatrix}}};
{\ar@3{->}(12,0);(-12,0)_{\begin{smallmatrix}m\\n\end{smallmatrix}}};
\endxy
\quad:=\quad
\xy0;/r.15pc/:
(25,0)*+{\mathcal{C}}="2";
(-25,0)*+{\mathcal{D}}="3";
{\ar@/_2.25pc/_{F} "2";"3"};
{\ar|-(0.5){G} "2";"3"};
{\ar@/^2.25pc/^{H} "2";"3"};
{\ar@2{->}(-8,12.5);(-8,1.5)_{\r}};
{\ar@2{->}(8,12.5);(8,1.5)^{\s}};
{\ar@3{->}(6,7);(-6,7)_{m}};
{\ar@{=>}(-8,-1.5);(-8,-12.5)_{\l}};
{\ar@{=>}(8,-1.5);(8,-12.5)^{\t}};
{\ar@3{->}(6,-7);(-6,-7)^{n}};
\endxy
\ee
and is defined by the assignment
\be
\xy0;/r.15pc/:
(-30,0)*+{x}="1";
(40,30)*+{F(x)}="2";
(40,0)*+{G(x)}="3";
(40,-30)*+{H(x)}="4";
{\ar@/^1.75pc/^{\s(x)} "2";"3"};
{\ar@/_1.75pc/_{\rho(x)} "2";"3"};
{\ar@/^1.75pc/^{\t(x)} "3";"4"};
{\ar@/_1.75pc/_{\l(x)} "3";"4"};
{\ar@2{->}(50,15);(30,15)_{m(x)}};
{\ar@2{->}(50,-15);(30,-15)_{n(x)}};
{\ar@{|->}^{\begin{smallmatrix}m\\ n\end{smallmatrix}} "1"+(20,0);(10,0)};
\endxy
\ee
for each object $x$ in $\mC.$ 

\bd
\label{defn:adjunction}
Let $\mC$ be a (strict) 2-category.
An \emph{\uline{adjunction}} in $\mC$
consists of a pair of objects $x,y$ in $\mC,$
a pair of morphisms 
\be
\xy0;/r.25pc/:
(-10,0)*+{x}="x";
(10,0)*+{y}="y";
{\ar@<1ex>"x";"y"^{f}};
{\ar@<1ex>"y";"x"^{g}};
\endxy
\ee
and a pair of 2-morphisms
\be
\xy0;/r.25pc/:
(-10,7.5)*+{x}="x1";
(10,7.5)*+{x}="x2";
(0,-7.5)*+{y}="y";
{\ar"x1";"x2"^{\id_{x}}};
{\ar"x1";"y"_{f}};
{\ar"y";"x2"_{g}};
{\ar@{=>}(0,7);"y"_(0.35){\h}};
\endxy
\aand
\xy0;/r.25pc/:
(-10,-7.5)*+{y}="y1";
(10,-7.5)*+{y}="y2";
(0,7.5)*+{x}="x";
{\ar"y1";"y2"_{\id_{y}}};
{\ar"y1";"x"^{g}};
{\ar"x";"y2"^{f}};
{\ar@{=>}"x";(0,-7)_(0.65){\e}};
\endxy
\ee
satisfying%
\be
\label{eq:zigzag1}
\xy0;/r.25pc/:
(-22.5,0)*+{x}="x1";
(-7.5,0)*+{y}="y1";
(7.5,0)*+{x}="x2";
(22.5,0)*+{y}="y2";
{\ar@/^2.25pc/"x1";"x2"^{\id_{x}}};
{\ar"x1";"y1"|-{f}};
{\ar"y1";"x2"|-{g}};
{\ar"x2";"y2"|-{f}};
{\ar@/_2.25pc/"y1";"y2"_{\id_{y}}};
{\ar@{=>}"y1"+(0,8.5);"y1"^(0.35){\h}};
{\ar@{=>}"x2";"x2"+(0,-8.5)^(0.55){\e}};
\endxy
\quad
=
\quad
\xy0;/r.25pc/:
(-10,0)*+{x}="x";
(10,0)*+{y}="y";
{\ar@/^1.5pc/"x";"y"^{f}};
{\ar@/_1.5pc/"x";"y"_{f}};
{\ar@{=>}(0,5);(0,-5)_{\id_{f}}};
\endxy
\ee
and
\be
\label{eq:zigzag2}
\xy0;/r.25pc/:
(-22.5,0)*+{y}="y1";
(-7.5,0)*+{x}="x1";
(7.5,0)*+{y}="y2";
(22.5,0)*+{x}="x2";
{\ar@/^2.25pc/"x1";"x2"^{\id_{x}}};
{\ar"y1";"x1"|-{g}};
{\ar"x1";"y2"|-{f}};
{\ar"y2";"x2"|-{g}};
{\ar@/_2.25pc/"y1";"y2"_{\id_{y}}};
{\ar@{=>}"y2"+(0,8.5);"y2"^(0.35){\h}};
{\ar@{=>}"x1";"x1"+(0,-8.5)^(0.55){\e}};
\endxy
\quad
=
\quad
\xy0;/r.25pc/:
(-10,0)*+{y}="y";
(10,0)*+{x}="x";
{\ar@/^1.5pc/"y";"x"^{g}};
{\ar@/_1.5pc/"y";"x"_{g}};
{\ar@{=>}(0,5);(0,-5)_{\id_{g}}};
\endxy
.
\ee
Conditions (\ref{eq:zigzag1}) and
(\ref{eq:zigzag2}) are known as the
\emph{\uline{zig-zag identities}}. 
An adjunction as above is typically
written as a quadruple $(f,g,\h,\e)$
and we say $f$ is \emph{\uline{left adjoint}} 
to $g$ and write $f\dashv g.$
\ed

A left adjoint is unique in the following sense. 

\blem
\label{lem:adjointunique}
Let $\mC$ be a (strict) 2-category and let
$x$ and $y$ be two objects in $\mC.$ 
Let $x\xleftarrow{g}y$ be a 1-morphism
and let 
$(f,g,\h,\e)$ and
$(f',g,\h',\e')$ be adjunctions 
in which which $f$ and $f'$ are both 
left adjoint to $g.$
Then there exists a vertically invertible
2-morphism
$\s:f\Rightarrow f'$ such that
\be
\label{eq:morphismadjunction1}
\xy0;/r.35pc/:
(-10,7.5)*+{x}="x1";
(10,7.5)*+{x}="x2";
(0,-7.5)*+{y}="y";
{\ar"x1";"x2"^{\id_{x}}};
{\ar"x1";"y"|-{f}};
{\ar"y";"x2"_{g}};
{\ar@/_2.25pc/"x1";"y"_{f'}};
{\ar@{=>}(0,7);"y"_(0.35){\h}};
{\ar@{=>}(-6,-0.5);(-10,-3)_(0.35){\s}};
\endxy
\quad
=
\quad
\xy0;/r.35pc/:
(-10,7.5)*+{x}="x1";
(10,7.5)*+{x}="x2";
(0,-7.5)*+{y}="y";
{\ar"x1";"x2"^{\id_{x}}};
{\ar"x1";"y"_{f'}};
{\ar"y";"x2"_{g}};
{\ar@{=>}(0,7);"y"_(0.35){\h'}};
\endxy
\ee
and
\be
\label{eq:morphismadjunction2}
\xy0;/r.35pc/:
(-10,-7.5)*+{y}="y1";
(10,-7.5)*+{y}="y2";
(0,7.5)*+{x}="x";
{\ar"y1";"y2"_{\id_{y}}};
{\ar"y1";"x"^{g}};
{\ar@/^2.25pc/"x";"y2"^{f}};
{\ar"x";"y2"|-{f'}};
{\ar@{=>}"x";(0,-7)_(0.65){\e'}};
{\ar@{=>}(10,3);(6,0.5)^{\s}};
\endxy
\quad
=
\quad
\xy0;/r.35pc/:
(-10,-7.5)*+{y}="y1";
(10,-7.5)*+{y}="y2";
(0,7.5)*+{x}="x";
{\ar"y1";"y2"_{\id_{y}}};
{\ar"y1";"x"^{g}};
{\ar"x";"y2"^{f}};
{\ar@{=>}"x";(0,-7)_(0.65){\e}};
\endxy
.
\ee
\elem

In this paper, we focus on an adjunction
in a particular 2-category obtained from
functors between 2-categories.

\bd
\label{defn:functor2cat}
Let  $\mC$ and $\mD$ be two (strict) 
2-categories. Let
$\mathbf{Fun}(\mC,\mathcal{D})$ 
be the 2-category whose objects
are (strict) functors from 
$\mC$ to $\mD,$ 1-morphisms are
oplax-natural transformations,
and 2-morphisms are modifications.
\ed

We spell out what it means to have
an adjunction in this 2-category
explicitly.

\blem
\label{lem:semipseudonaturaladjunction}
Let $\mathcal{C}$ and $\mathcal{D}$ be two 
(strict)
2-categories and let 
$\mathbf{Fun}(\mC,\mathcal{D})$ 
be the functor 2-category described in
Definition (\ref{defn:functor2cat}).
An adjunction in $\mathbf{Fun}(\mC,\mathcal{D})$
consists of two (strict) functors
$F,G:\mathcal{C}\to\mathcal{D},$
two oplax-natural transformations 
$\s:F\Rightarrow G$ and $\rho:G\Rightarrow F,$ 
and two modifications 
$\h:\id_{F} \Rrightarrow\begin{smallmatrix}\s \\
\rho\end{smallmatrix}$ 
and
$\e:\begin{smallmatrix}\rho \\ \s\end{smallmatrix}
\Rrightarrow \id_{G}$ 
such that the diagrams
\be
\xy 0;/r.25pc/:
(10,-7.5)*+{\begin{smallmatrix}
\rho\\ \id_{F}\end{smallmatrix}}="1";
(0,7.5)*+{\begin{smallmatrix}
\rho\\ \s\\ \rho\end{smallmatrix}}="2";
(-10,-7.5)*+{\begin{smallmatrix}
\id_{G}\\ \rho\end{smallmatrix}}="3";
{\ar@3{->}"1";"2"_{\begin{smallmatrix}
\id_{\rho}\\ \h\end{smallmatrix}}};
{\ar@3{->}"2";"3"_{\begin{smallmatrix}
\e\\ \id_{\rho}\end{smallmatrix}}};
{\ar@3{->}"1";"3"^{\id_{\rho}}};
\endxy
\qquad \& \qquad 
\xy0;/r.25pc/:
(-10,-7.5)*+{\begin{smallmatrix}
\id_{F}\\ \s\end{smallmatrix}}="1";
(0,7.5)*+{\begin{smallmatrix}
\s\\ \rho\\ \s\end{smallmatrix}}="2";
(10,-7.5)*+{\begin{smallmatrix}
\s\\ \id_{G}\end{smallmatrix}}="3";
{\ar@3{->}"1";"2"^{\begin{smallmatrix}
\h\\ \id_{\s}\end{smallmatrix}}};
{\ar@3{->}"2";"3"^{\begin{smallmatrix}
\id_{\s}\\ \e\end{smallmatrix}}};
{\ar@3{->}"1";"3"_{\id_{\s}}};
\endxy
\ee
both commute, i.e.
\be
\xy0;/r.25pc/:
(0,22.5)*+{G}="1";
(0,7.5)*+{F}="2";
(0,-7.5)*+{G}="3";
(0,-22.5)*+{F}="4";
{\ar@2{->}"1";"2"_{\rho}};
{\ar@2{->}"2";"3"_{\s}};
{\ar@2{->}"3";"4"_{\rho}};
{\ar@2{->}@/_2.25pc/"2";"4"_{\id_{F}}};
{\ar@2{->}@/^2.25pc/"1";"3"^{\id_{G}}};
{\ar@3{->}(-7.5,-7.5);"3"^(0.3){\h}};
{\ar@3{->}"2";(7.5,7.5)^(0.6){\e}};
\endxy
=
\quad
\xy0;/r.25pc/:
(0,10)*+{G}="1";
(0,-10)*+{F}="2";
{\ar@2{->}@/_1.50pc/"1";"2"_{\rho}};
{\ar@2{->}@/^1.50pc/"1";"2"^{\rho}};
{\ar@3{->}(-5,0);(5,0)^{\id_{\rho}}};
\endxy
\quad
\&
\xy0;/r.25pc/:
(0,22.5)*+{F}="1";
(0,7.5)*+{G}="2";
(0,-7.5)*+{F}="3";
(0,-22.5)*+{G}="4";
{\ar@2{->}"1";"2"_{\s}};
{\ar@2{->}"2";"3"_{\rho}};
{\ar@2{->}"3";"4"_{\s}};
{\ar@2{->}@/^2.25pc/"2";"4"^{\id_{G}}};
{\ar@2{->}@/_2.25pc/"1";"3"_{\id_{F}}};
{\ar@3{->}(-7.5,7.5);"2"^(0.3){\h}};
{\ar@3{->}"3";(7.5,-7.5)^(0.6){\e}};
\endxy
=
\quad
\xy0;/r.25pc/:
(0,10)*+{F}="1";
(0,-10)*+{G}="2";
{\ar@2{->}@/_1.50pc/"1";"2"_{\s}};
{\ar@2{->}@/^1.50pc/"1";"2"^{\s}};
{\ar@3{->}(-5,0);(5,0)^{\id_{\s}}};
\endxy
,
\ee
respectively. 
\elem

\br
\label{rmk:zigzag}
Because the zig-zag identities only involve
the equality of modifications, and since
the datum of a modification
consists only of an assignment of
2-morphisms in $\mathcal{D}$ to objects
of $\mathcal{C},$ they can be re-expressed as
\be
\xy0;/r.25pc/:
(0,22.5)*+{G(x)}="1";
(0,7.5)*+{F(x)}="2";
(0,-7.5)*+{G(x)}="3";
(0,-22.5)*+{F(x)}="4";
{\ar"1";"2"|-{\rho(x)}};
{\ar"2";"3"|-{\s(x)}};
{\ar"3";"4"|-{\rho(x)}};
{\ar@/_2.75pc/"2";"4"_{\id_{F(x)}}};
{\ar@/^2.75pc/"1";"3"^{\id_{G(x)}}};
{\ar@2{->}(-10,-7.5);"3"^(0.3){\h(x)}};
{\ar@2{->}"2";(10,7.5)^(0.7){\e(x)}};
\endxy
=
\quad
\xy0;/r.25pc/:
(0,10)*+{G(x)}="1";
(0,-10)*+{F(x)}="2";
{\ar@/_1.50pc/"1";"2"_{\rho(x)}};
{\ar@/^1.50pc/"1";"2"^{\rho(x)}};
{\ar@2{->}(-5,0);(5,0)^{\id_{\rho(x)}}};
\endxy
\ee
and
\be
\xy0;/r.25pc/:
(0,22.5)*+{F(x)}="1";
(0,7.5)*+{G(x)}="2";
(0,-7.5)*+{F(x)}="3";
(0,-22.5)*+{G(x)}="4";
{\ar"1";"2"|-{\s(x)}};
{\ar"2";"3"|-{\rho(x)}};
{\ar"3";"4"|-{\s(x)}};
{\ar@/^2.75pc/"2";"4"^{\id_{G(x)}}};
{\ar@/_2.75pc/"1";"3"_{\id_{F(x)}}};
{\ar@2{->}(-10,7.5);"2"^(0.3){\h(x)}};
{\ar@2{->}"3";(10,-7.5)^(0.7){\e(x)}};
\endxy
=
\quad
\xy0;/r.25pc/:
(0,10)*+{F(x)}="1";
(0,-10)*+{G(x)}="2";
{\ar@/_1.50pc/"1";"2"_{\s(x)}};
{\ar@/^1.50pc/"1";"2"^{\s(x)}};
{\ar@2{->}(-5,0);(5,0)^{\id_{\s(x)}}};
\endxy
\ee
for every object $x$ of $\mC,$ i.e. for every object $x$ in $\mC,$
the quadruple $\big(\s(x),\rho(x),\h(x),\e(x)\big)$
is an adjunction.
\er

\section*{Acknowledgments}

I would like to thank V. P. Nair for helpful conversations
and for bringing my attention to \cite{BGdQRL},
which was the seed of the ideas in this work.
Amol Deshmukh and Dennis Sullivan 
convinced me to read \cite{Ja1}, which
has offered additional insight to these ideas.
I would also like to thank Masoud Khalkhali for
bringing a few errors to my attention
and Scott O. Wilson for several helpful
suggestions.
Finally, I am grateful to an anonymous referee 
of Applied Categorical Structures for their valuable comments.
The majority of this work was completed when the author was
at the City College of New York and the CUNY Graduate Center
and was part of the author's Ph.D. thesis \cite{Pa16}.
Work by this author was partially supported 
by the CUNY Graduate Center Capelloni
Dissertation Fellowship and NSF grant PHY-1213380.

\section*{Index of notation}

\begin{longtable}{c|c|c|c}
\hline
Notation & Name/description & Location & Page \\
\hline
$\mA$ & unital $C^*$-algebra & Def'n \ref{defn:C*algebra}
& \pageref{defn:C*algebra}\\
\hline
$\CAlg$ & category of unital $C^*$-algebras 
& Def'n \ref{defn:CAlg} & \pageref{defn:CAlg}\\
\hline
$\w$ & a state (on some $C^*$-algebra) & Def'n \ref{defn:state} & \pageref{defn:state}\\
\hline
$\mS(\mA)$ & set of states on $\mA$ 
& Def'n \ref{defn:state} & \pageref{defn:state}\\
\hline
$\Rep(\mA)$ & category of representations of $\mA$
& Def'n \ref{defn:RepA} & \pageref{defn:RepA}\\
\hline
$\Hi$ & Hilbert space
& Def'n \ref{defn:RepA} & \pageref{defn:RepA}\\
\hline
$\mB(\Hi)$ & bounded linear operators on $\Hi$
& Def'n \ref{defn:RepA} & \pageref{defn:RepA}\\
\hline
$\mS$ &states pre-sheaf 
&Prop \ref{prop:statefunctor}&\pageref{prop:statefunctor}\\
\hline
$\St$ &states pre-stack
&Eqn (\ref{eq:statesfunctor}) &\pageref{eq:statesfunctor}\\
\hline
$\Rep$ & representation pre-stack 
&Eqn (\ref{eq:Repstack}) & \pageref{eq:Repstack}\\
\hline
$\mathcal{N}_{\w}$ & null-space associated to $\w$
& Eqn (\ref{eq:nullspace}) &\pageref{eq:nullspace}\\
\hline
$[a]$ & typical element of $\mA/\mathcal{N}_{\w}$
& Eqn (\ref{eq:GNSinnerproduct}) 
&\pageref{eq:GNSinnerproduct}\\
\hline
$\Hi_{\w}$ & Hilbert space associated to $\w$ via GNS
& Eqn (\ref{eq:GNSHilbertspace}) 
&\pageref{eq:GNSHilbertspace}\\
\hline
$\pi_{\w}$ & representation associated to $\w$ via GNS
& Eqn (\ref{eq:GNSrepn}) &\pageref{eq:GNSrepn}\\
\hline
$\GNS_{\mA}$ & GNS construction for $\mA$
& Con \ref{con:GNSobjects} 
&\pageref{con:GNSobjects}--\pageref{eq:GNSrepn}\\
\hline
$\GNS_{f}$ & GNS construction for 
$\mA'\xrightarrow{f}\mA$
& Con \ref{con:GNSmorphisms} 
&\pageref{con:GNSmorphisms}--\pageref{eq:GNSf} \\
\hline
$\GNS$ & the GNS construction 
&Thm \ref{thm:GNSfunctor} &\pageref{thm:GNSfunctor}\\
\hline
$\W$ & unit vector (occasionally cyclic) 
&Def'n \ref{defn:cyclic}&\pageref{defn:cyclic}\\
\hline
$(\pi,\Hi,\W)$ & pointed (or cyclic) representation 
&Def'n \ref{defn:cyclic}&\pageref{defn:cyclic}\\
\hline
$\pRep(\mA)$ & category of pointed
representations of $\mA$
&Def'n \ref{defn:pointedRepobjects}
&\pageref{defn:pointedRepobjects}\\
\hline
$\cRep(\mA)$
&category of cyclic representations of $\mA$
&Def'n \ref{defn:pointedRepobjects}
&\pageref{defn:pointedRepobjects}\\
\hline
$\rest_{\mA}$ & restriction to states on $\mA$ functor
&Con \ref{con:restrictionfunctor}
&\pageref{con:restrictionfunctor}\\
\hline
$\w_{\W}$ & vector state $\<\W,\;\cdot\;\W\>$
&Eqn \ref{eq:omegaOmega}
&\pageref{eq:omegaOmega}\\
\hline
$\pRep$
&pointed representation pre-stack
& Con \ref{con:pointedRepmorphisms}
&\pageref{con:pointedRepmorphisms}\\
\hline
$\rest_{f}$ & restriction to states for 
$\mA'\xrightarrow{f}\mA$
& Con \ref{con:pointedRepmorphisms}
&\pageref{con:pointedRepmorphisms}\\
\hline
$\rest$ & restriction natural transformation
& Prop \ref{prop:rest}
&\pageref{prop:rest}\\
\hline
$\pGNS$
&pointed GNS construction
& Thm \ref{thm:pointedGNS}
&\pageref{thm:pointedGNS}\\
\hline
$m$ &GNS modification &Lem \ref{lem:m}
&\pageref{lem:m}\\
\hline
\end{longtable}

\bibliographystyle{utcaps}
\bibliography{GNSadjunction}

\providecommand{\href}[2]{#2}\begingroup\raggedright\begin{thebibliography}{10}

\bibitem{GN43}
I.~Gelfand and M.~Neumark, ``On the imbedding of normed rings into the ring of
  operators in {H}ilbert space,'' {\em Rec. Math. [Mat. Sbornik] N.S.}
  {\bfseries 12(54)} (1943) 197--213.

\bibitem{Se47}
I.~E. Segal, ``Irreducible representations of operator algebras,'' {\em Bull.
  Amer. Math. Soc.} {\bfseries 53} (1947) 73--88.

\bibitem{Ma98}
S.~Mac~Lane, {\em Categories for the working mathematician}, vol.~5 of {\em
  Graduate Texts in Mathematics}.
\newblock Springer-Verlag, New York, second~ed., 1998.

\bibitem{Di77}
J.~Dixmier, {\em {$C\sp*$}-algebras}.
\newblock North-Holland Publishing Co., Amsterdam-New York-Oxford, 1977.
\newblock Translated from the French by Francis Jellett, North-Holland
  Mathematical Library, Vol. 15.

\bibitem{Fi96}
P.~A. Fillmore, {\em A user's guide to operator algebras}.
\newblock Canadian Mathematical Society Series of Monographs and Advanced
  Texts. John Wiley \& Sons, Inc., New York, 1996.
\newblock A Wiley-Interscience Publication.

\bibitem{HaQM}
B.~C. Hall, \href{http://dx.doi.org/10.1007/978-1-4614-7116-5}{{\em Quantum
  theory for mathematicians}}, vol.~267 of {\em Graduate Texts in Mathematics}.
\newblock Springer, New York, 2013.

\bibitem{Ja1}
E.~T. Jaynes, ``{Information Theory and Statistical Mechanics},''
  \href{http://dx.doi.org/10.1103/PhysRev.106.620}{{\em Phys. Rev.} {\bfseries
  106} (1957) 620--630}.

\bibitem{Ha15}
B.~C. Hall, \href{http://dx.doi.org/10.1007/978-3-319-13467-3}{{\em Lie groups,
  {L}ie algebras, and representations}}, vol.~222 of {\em Graduate Texts in
  Mathematics}.
\newblock Springer, second~ed., 2015.

\bibitem{BGdQRL}
A.~P. Balachandran, T.~R. Govindarajan, A.~R. de~Quieroz, and A.~F. Reyes-Lega,
  ``Entanglement, Particle Identity and the GNS Construction: A Unifying
  Approach,'' \href{http://dx.doi.org/10.1103/PhysRevLett.110.080503}{{\em
  Phys. Rev. Lett.} {\bfseries 110} (2013) 080503},
  \href{http://arxiv.org/abs/1303.0688}{{\ttfamily arXiv:1303.0688 [hep-th]}}.

\bibitem{Wa94}
R.~M. Wald, {\em Quantum field theory in curved spacetime and black hole
  thermodynamics}.
\newblock Chicago Lectures in Physics. University of Chicago Press, Chicago,
  IL, 1994.

\bibitem{Br93}
J.-L. Brylinski, \href{http://dx.doi.org/10.1007/978-0-8176-4731-5}{{\em Loop
  spaces, characteristic classes and geometric quantization}}.
\newblock Modern Birkh\"auser Classics. Birkh\"auser Boston, Inc., Boston, MA,
  2008.
\newblock Reprint of the 1993 edition.

\bibitem{Fo08}
G.~B. Folland, \href{http://dx.doi.org/10.1090/surv/149}{{\em Quantum field
  theory: A tourist guide for mathematicians}}, vol.~149 of {\em Mathematical
  Surveys and Monographs}.
\newblock American Mathematical Society, Providence, RI, 2008.

\bibitem{Be}
J.~B{\'e}nabou, \href{http://dx.doi.org/10.1007/BFb0074299}{``Introduction to
  bicategories,''} in {\em Reports of the Midwest Category Seminar}, pp.~1--77.
\newblock Springer Berlin Heidelberg, Berlin, Heidelberg, 1967.

\bibitem{Bo94}
F.~Borceux, {\em Handbook of categorical algebra. 1}, vol.~50 of {\em
  Encyclopedia of Mathematics and its Applications}.
\newblock Cambridge University Press, Cambridge, 1994.

\bibitem{BD}
J.~C. Baez and J.~Dolan, ``From finite sets to {F}eynman diagrams,'' in {\em
  Mathematics unlimited---2001 and beyond}, pp.~29--50.
\newblock Springer, Berlin, 2001.
\newblock \href{http://arxiv.org/abs/math/0004133}{{\ttfamily math/0004133}}.

\bibitem{Pa16}
A.~Parzygnat, ``Some 2-Categorical Aspects in Physics,'' 2016.
\newblock \url{http://academicworks.cuny.edu/gc_etds/1475}. Ph.D. Thesis {\it
  CUNY Academic Works.}

\end{thebibliography}\endgroup

\end{document}